\documentclass[twocolumn,aps,superscriptaddress,longbibliography,amsmath,amssymb,reprint]{revtex4-2} 
\usepackage{bm}
\usepackage{amsmath}
\usepackage{amssymb}
\usepackage{times}
\usepackage[usenames,dvipsnames]{color}
\usepackage[table]{xcolor}
\usepackage{graphicx}
\usepackage{dcolumn}
\usepackage{simplewick}
\usepackage{hyperref}
\usepackage{romannum}
\usepackage{dcolumn}
\usepackage[utf8]{inputenc}
\usepackage[T1]{fontenc}
\usepackage{mathptmx}
\DeclareUnicodeCharacter{2010}{-}

\hypersetup{
  colorlinks=false,
  colorlinks=true,
  citecolor=blue,
  linkcolor=blue,
  urlcolor=blue}

\begin{document}

\title{Probing transition rates, nuclear moments and electric dipole polarizability 
in nobelium using multireference FSRCC and PRCC theories}

\author{Ravi Kumar}
\email{Authors contributed equally to this work}
\affiliation{Department of Energy and Materials, Faculty of Science and Engineering, 
             Kindai University, Osaka 577-8502, Japan}

\author{Palki Gakkhar}
\email{Authors contributed equally to this work}
\affiliation{Department of Physics, Indian Institute of Technology, 
             Hauz Khas, New Delhi 110016, India}

\author{D. Angom}
\affiliation{Department of Physics, Manipur University,
             Canchipur 795003, Manipur, India}

\author{B. K. Mani}
\email{bkmani@physics.iitd.ac.in}
\affiliation{Department of Physics, Indian Institute of Technology, 
             Hauz Khas, New Delhi 110016, India}

\begin{abstract}
We employ an all-particle multireference Fock-space relativistic 
coupled-cluster (FSRCC) theory to compute the ionization potential, 
excitation energy, transition rate and hyperfine structure constants associated 
with $7s^2\;^{1}S_{0}\rightarrow 7s7p\;^{3}P_{1}$ 
and $7s^2\;^{1}S_{0}\rightarrow 7s7p\;^1P_{1}$ transitions in nobelium (No). 
Using our state-of-the-art calculations in conjunction with available 
experimental data [S. Raeder {\em et al.}, Phys. Rev. Lett. 120, 232503 (2018)], 
we extract the values of nuclear 
magnetic dipole ($\mu$) and electric quadrupole ($Q$) moments for $^{253}$No.
Further, information on nuclear deformation in even-mass isotopes is 
extracted from the isotope shift calculations. Moreover, we employ a
perturbed relativistic coupled-cluster (PRCC) theory to compute the 
ground state electric dipole polarizability of No. In addition, 
to assess the accuracy of our calculations, we compute the ionization 
potential and dipole polarizability of lighter homolog ytterbium (Yb).
To account for strong relativistic and quantum electrodynamical 
(QED) effects in No, we incorporate the corrections from Breit 
interaction, vacuum polarization and self-energy in our calculations. 
The contributions from triple excitations in coupled-cluster is 
accounted perturbatively.
Our calculations reveal a significant contribution of $\approx$10\% 
from the perturbative triples to the transition rate of 
$7s^2\;^1S_{0}\rightarrow 7s7p\;^3P_{1}$ transition. The largest 
cumulative contribution from Breit+QED is observed to be $\approx$4\%,
to the magnetic dipole hyperfine structure constant 
of $7s7p\;^1P_{1}$ state. 
Our study provides a comprehensive understanding of atomic and nuclear 
properties of nobelium with valuable insights into the electron correlation 
and relativistic effects in superheavy elements.
\end{abstract}
\maketitle

\section{Introduction}

The study of atomic, nuclear, and chemical properties of superheavy elements
(SHEs) is an area of significant scientific 
interests \cite{turler-13, schadel-14, pershina-15, peter-15, eliav-15, giuliani-19}.
However, due to their
extremely low production rates, often as low as few atoms per second at most, and
short half-lives, experimental investigation of their properties 
is nontrivial \cite{turler-13, schadel-14, hoffman-08}.
The specialized facilities required to process single-atom-at-a-time
restrict direct measurements. Considering this, an effective
approach for studying SHEs is through the high precision atomic 
structure and properties calculations. Atomic structure calculations can play a vital 
role in identifying the atomic levels, probing ground and excited state 
properties, and exploring the nuclear characteristics of 
SHEs \cite{laatiaoui-16, raeder-18}.
This, however, is also a challenging task as SHEs exhibit strong relativistic 
and QED effects due to their high nuclear charge \cite{eliav-15}. 
These effects modify orbital energy levels leading to shifts in the 
ground and excited state electron configurations.
For a reliable prediction of the properties of SHEs using precision structure
calculations, both relativistic and correlation effects should be treated at
the highest level of accuracy. Moreover, it is also essential to employ large
basis sets in the calculations to ensure the convergence of the properties.

Among SHEs, nobelium (Z = 102) has received a special attention due to recent 
spectroscopic measurements \cite{laatiaoui-16, raeder-18, chhetri-18}. 
Notably, it is the only transfermium element for which hyperfine spectra and 
isotope shifts have been measured using laser spectroscopy experiments \cite{raeder-18}. 
The first breakthrough in No came in 2016 when Laatiaoui {\em et al.} \cite{laatiaoui-16} 
successfully identified the $7s^2{\;^1S_0} \rightarrow 7s7p{\;^1P_1}$ transition 
in $^{254}$No and measured its ionization potential and transition rate 
using a single-atom-at-a-time experiment. This marked the first optical
spectroscopic study in transfermium elements. 
Later, in 2018, an improved technique allowed a more accurate measurement 
of ionization potential with an uncertainty of 50 $\mu$eV \cite{chhetri-18}. 
In the same year, Raeder {\em et al.} \cite{raeder-18} measured the hyperfine 
spectra of $^{253}$No and the isotope shifts of $^{252}$No and $^{253}$No 
relative to $^{254}$No. And most recently, in 2024, the isotope shift 
of $^{255}$No relative to $^{254}$No was measured by 
Warbinek {\em et al.} \cite{warbinek-24}.

The experimental advancements in No has established it as a benchmark 
superheavy candidate for testing the state-of-the-art relativistic 
many-body methods. Accurate theoretical predictions become more critical 
for excited states and related properties where experimental data is 
often scarce, and electron correlation and relativistic effects are 
highly complex. 
In addition, the multireference nature of the states in No puts further 
hurdles in terms of defining the model wavefunction and the divergence 
due to intruder states. 
At present, theoretical investigations of excited state properties of No
are limited to few calculations \cite{liu-07, borschevsky-07, indelicato-07}. 
There is a large variation in the excited state properties reported in
these works. For instance, Refs. \cite{liu-07, indelicato-07} use 
multiconfiguration Dirac-Fock (MCDF) theory to compute the transition 
rate of $^1S_{0} \rightarrow {^1P_{1}}$ transition. Though the same 
theory is used in both the works, value of transition rate reported 
in Ref. \cite{liu-07} is $\approx$ 29\% higher than that in Ref. \cite{indelicato-07}. 
The reason for this could be the inherent dependencies of results on 
the choice of configurations to incorporate electron correlation effects 
in this theory.
The third result is using the relativistic configuration interaction (RCI) 
method \cite{borschevsky-07} and is higher than MCDF 
values \cite{liu-07, indelicato-07}. Considering this, it can thus be
surmised that there is a gap in terms of the availability of accurate 
theory data on the properties of No. One of the main aims of 
the present work is to fill this gap.

In this work, we employ an all-particle FSRCC
theory \cite{mani-11,ravi-21a,palki-24} for two-valence systems to compute 
ionization potential (IP), transition rate,
and hyperfine structure constants associated with
$7s^2\;{^1}S_0 \rightarrow 7s7p\;{^1}P_1$ and $7s^2\;{^1}S_0 \rightarrow 7s7p\;{^3}P_1$
transitions in No.
The hyperfine constants are used further to extract the nuclear magnetic 
dipole ($\mu$) and electric quadrupole ($Q$) moments. Moreover, to investigate 
nuclear deformation
of even-mass isotopes, we have performed isotope shift calculations using
multiconfiguration Dirac-Fock (MCDF) method, results from which are used
further to extract the mean square charge radii of the isotopes
of nobelium. Furthermore, we employ a perturbed relativistic coupled-cluster
(PRCC) theory \cite{chattopadhyay-12b, chattopadhyay-13b, chattopadhyay-14, 
chattopadhyay-15, ravi-20} to compute the electric dipole polarizability ($\alpha$) of the ground
state of No. The dipole polarizability of an atom or ion is a fundamental property 
that quantifies how easily its electron cloud distorts in response to an external 
electric field. In superheavy elements, strong relativistic effects significantly 
alter both the inner and outer core orbitals' structure, leading to a 
pronounced influence on the response of the electron cloud \cite{eliav-15}. 
Therefore, studying the electric dipole polarizability of SHEs provides 
valuable insight into the role of relativistic effects in determining 
atomic response properties \cite{thierfelder-09}.

In addition, to assess the accuracy of our results, we calculate
the ionization potential and $\alpha$ for well-studied
homolog ytterbium (Yb).
The FSRCC method employed in present work to calculate the excited
state transition properties of No is one of the most accurate many-body
methods for atomic structure and properties calculations as it accounts for electron
correlation to all orders of residual Coulomb interaction. Similarly, the
PRCC theory used to calculate $\alpha$ does not employ a sum-over states
approach \cite{safronova-99, derevianko-99}, and therefore accounts 
for external perturbation
more accurately. It has been successfully applied to calculate $\alpha$
for several atoms and ions \cite{chattopadhyay-12b,
chattopadhyay-13b, chattopadhyay-14, chattopadhyay-15, ravi-20}.
In addition, to improve the accuracy of our results further, we also
incorporate the corrections from the Breit interaction, QED effects,
and perturbative triples in our calculations.

The remainder of the paper is organized into four sections.
In Sec. II, we provide a brief discussion on the FSRCC and PRCC
theories. In Sec. III, we present and discuss our results of
ionization potential, transition rate, hyperfine structure
constants and nuclear moments, isotope shift, and electric
dipole polarizability in different subsections. In Sec. IV,
the theoretical uncertainty in our calculation is discussed.
Unless stated otherwise, all the results presented in this
paper are in atomic units ( $\hbar=m_e=e=1/4\pi\epsilon_0=1$).

\section{Methodology}

For the calculation of ionization potentials, excitation energies, E1 transition 
amplitudes and hyperfine structure constants we have used a two-valence FSRCC theory. 
The details related to the implementation of this theory is provided in our 
previous works \cite{mani-11, ravi-21a}. In addition, the calculation of dipole 
polarizability requires an atomic many-body theory which can account for 
external perturbations accurately in the calculation. For this, we used PRCC 
theory developed in our previous 
works \cite{mani-09, chattopadhyay-12a, chattopadhyay-14, ravi-20, ravi-21b}. 
For completeness, below we provide a very brief description of these theories. 

\subsection{Two-valence FSRCC theory}

The many-body wavefunction for 
a two-valence atom or ion is obtained by solving the eigenvalue equation
\begin{equation}
  H^{\rm DCB}|\Psi_{vw} \rangle = E_{vw} |\Psi_{vw} \rangle,
  \label{hdc_2v}
\end{equation}
where $E_{vw}$ is the exact energy. And, $H^{\rm DCB}$ is the 
Dirac-Coulomb-Breit no-virtual-pair Hamiltonian, expressed as 
\begin{eqnarray}
   H^{\rm DCB} & = & \sum_{i=1}^N \left [c\bm{\alpha}_i \cdot
        \mathbf{p}_i + (\beta_i -1)c^2 - V_{N}(r_i) \right ]
                       \nonumber \\
    & & + \sum_{i<j}\left [ \frac{1}{r_{ij}}  + g^{\rm B}(r_{ij}) \right ].
  \label{ham_dcb}
\end{eqnarray}
Here, $\bm{\alpha}$ and $\beta$ are the Dirac matrices, and the last two 
terms, $1/r_{ij} $ and $g^{\rm B}(r_{ij})$, represent the Coulomb and Breit 
interactions, respectively.

In FSRCC, $|\Psi_{vw} \rangle$ is expressed in terms of the excitation 
operators as
\begin{equation}
	|\Psi_{vw}\rangle = e^{T^{(0)}} \left[ 1 + S^{(0)}_1 + S^{(0)}_2 + 
	\frac{1}{2} \left({S^{(0)}_1}^2 + {S^{(0)}_2}^2 \right) + 
	R^{(0)}\right ]|\Phi_{vw}\rangle,
  \label{2v_exact}
\end{equation}
where $v, w, \ldots$ represent the valence orbitals and 
$|\Phi_{vw}\rangle$ is the Dirac-Fock reference state. $|\Phi_{vw}\rangle$ is 
obtained by adding two electrons to the Dirac-Fock state for closed-shell 
configuration, $a^\dagger_wa^\dagger_v |\Phi_0\rangle$. The excitation operators 
$T^{(0)}$, $S^{(0)}$ and $R^{(0)}$ are referred to as the coupled-cluster (CC) 
operators for closed-shell, one-valence and two-valence sectors, respectively,
of the Hilbert space of all electrons.
For an atomic system with $N$-electrons, $T^{(0)}$, $S^{(0)}$ and $R^{(0)}$ 
operators in principle can have all possible excitations, and therefore, 
can be written as
\begin{equation}
	T^{(0)} = \sum_{i=1}^{N-2} T^{(0)}_i, \;\; 
	S^{(0)}= \sum_{i=1}^{N-1} S^{(0)}_i, \;\; {\rm and} \;\; 
	R^{(0)}= \sum^N_{i=1} R^{(0)}_i.
\end{equation}
Since residual Coulomb interaction is a two-body operator, the single and 
double excitations subsume most of the electron correlation effects 
and provide a good description of the atomic properties. 
We can, therefore, approximate 
$T^{(0)} = T^{(0)}_1 + T^{(0)}_2$, $S^{(0)} = S^{(0)}_1 + S^{(0)}_2$ 
and $R^{(0)} = R^{(0)}_2$. The CC theory with this approximation is 
referred to as the coupled-cluster with singles and doubles (CCSD) 
approximation. These one- and two-body CC operators can further be 
expressed in terms of electron creation and annihilation 
operators, as 
\begin{subequations}
\begin{eqnarray}
   T^{(0)}_1  = \sum_{ap}t_a^p a_p^{\dagger}a_a {\;\; \rm and \;\;} 
   T^{(0)}_2  = \frac{1}{2!}\sum_{abpq}t_{ab}^{pq} a_p^{\dagger}a_q^{\dagger}a_ba_a,
\end{eqnarray}
\begin{eqnarray}
   S^{(0)}_1 = \sum_{p}s_v^p a_p^{\dagger}a_v  {\;\; \rm and \;\;}
	S^{(0)}_2 = \sum_{apq}s_{va}^{pq} a_p^{\dagger}a_q^{\dagger}a_aa_v,
\end{eqnarray}
\begin{eqnarray}
	R^{(0)}_2 = \sum_{pq}r_{vw}^{pq} a_p^{\dagger}a_q^{\dagger}a_wa_v.
\end{eqnarray}
 \label{t1t2}
\end{subequations}
Here, the indices $a, b, \ldots$, $v, w, \ldots$ and $p, q, \ldots$ represent 
the core, valence and virtual orbitals, respectively. 
And, $t_{\ldots}^{\ldots}$, $s_{\ldots}^{\ldots}$ and $r_{\ldots}^{\ldots}$
represent the cluster amplitudes corresponding to $T$, $S$ and $R$ 
operators, respectively.

The operators $T^{(0)}$ and $S^{(0)}$ are obtained by solving the set of 
coupled nonlinear equations for closed-shell \cite{mani-09} and 
one-valence \cite{mani-10} sectors, respectively. The details related to
the computational implementation of RCC for closed-shell and one-valence 
systems in the form of a Fortran code is given in Ref. \cite{mani-17}. 
The operator $R^{(0)}_2$ is obtained by solving the CC 
equation for two-valence \cite{mani-11, ravi-21a, palki-24}
\begin{eqnarray}
   \langle\Phi^{pq}_{vw}|
    \bar H_{\rm N} +
   \{\contraction{\bar}{H}{_{\rm N}}{S}\bar H_{\rm N}S^{'}\} +
   \{\contraction{\bar}{H}{_{\rm N}}{S}\bar H_{\rm N}R^{(0)}_2\}
   |\Phi_{vw}\rangle = \nonumber \\
   E^{\rm att}_{vw}
   \langle\Phi^{pq}_{vw}|\Bigl[S^{'} + R^{(0)}_2 \Bigr]|\Phi_{vw}\rangle.
   \label{ccsd_2v}
\end{eqnarray}
Here, for compact notation we have used
$S' = S^{(0)}_1 + S^{(0)}_2 + \frac{1}{2}({S_1^{(0)}}^2 + {S_2^{(0)}}^2)$. 
The parameter $E^{\rm att}_{vw}$ on the right hand side of the equation is 
two-electron attachment energy, expressed as
\begin{equation}
  E^{\rm att}_{vw} = \epsilon_v + \epsilon_w + \Delta E^{\rm att}_{vw},
  \label{2v_eatt}
\end{equation}
where $\epsilon_v$ and $\epsilon_w$ are the Dirac-Fock energies of the 
valence electrons in $|\phi_v\rangle$ and $|\phi_w\rangle$ states, respectively. 
And, $\Delta E^{\rm att}_{vw},  = \Delta E^{\rm corr}_{vw} -
\Delta E^{\rm corr}_0$, is the difference of electron correlation energies 
of closed-shell and two-valence sectors.
And, $\bar{H}_N, = e^{-T^{(0)}} H_N e^{T^{(0)}}$, is a similarity
transformed Hamiltonian, which using Wick's theorem, can be reduced 
to the form
\begin{eqnarray}
  \bar{H}_N &=& H_N + \{\contraction[0.4ex]{} {H}{_N} {T} H_N T^{(0)}\} 
          + \frac{1}{2!} \{\contraction[0.4ex]{}{H}{_N}{T} 
         \contraction[0.8ex] {}{H}{_NT_0}{T^{(0)}}H_NT^{(0)}T^{(0)} \} 
                  +\nonumber \\ 
           && \frac{1}{3!}\{\contraction[0.4ex]{}{H}{_N}{T} 
          \contraction[0.8ex] {}{H}{_N T_0}{T^{(0)}}
          \contraction[1.2ex]{}{H}{_N T^{(0)} T_0}{T^{(0)}} 
          H_N T^{(0)} T^{(0)} T^{(0)} \} 
          + \frac{1}{4!}\{\contraction[0.4ex]{}{H}{_N}{T} 
          \contraction[0.8ex] {}{H}{_N T_0}{T^{(0)}}
          \contraction[1.2ex]{}{H}{_N T^{(0)} T_0}{T^{(0)}} 
          \contraction[1.6ex]{}{H}{_N T^{(0)}T^{(0)} T_0}{T^{(0)}} 
          H_N T^{(0)} T^{(0)} T^{(0)} T^{(0)} \}. 
\label{hnbar}   
\end{eqnarray}

\subsection{PRCC theory and electric dipole polarizability}

When an external electric field is applied to an atom or an ion, 
it modifies the wavefunctions of the system. We refer these modified 
wavefunctions as the perturbed wavefunctions, and for ground state 
we can denote it as $|\widetilde{\Psi}_0\rangle$. 
In PRCC theory, $|\widetilde{\Psi}_0\rangle$ is expressed as
\begin{equation}
 |\widetilde{\Psi}_0\rangle = e^{T^{(0)}}\left [ 1 + \lambda
  \mathbf{T}^{(1)}\cdot \mathbf{E}_{\rm ext} \right ] |\Phi_0\rangle,
 \label{psi0-ptrb}
\end{equation}
where $\mathbf{E}_{\rm ext}$ is an external electric field, the operator 
$\mathbf{T}^{(1)}$ is referred to as the perturbed CC operator 
and $\lambda$ is a perturbation parameter. The perturbed wavefunction
is an eigenstate of the modified Hamiltonian 
$H_{\rm Tot} =  H^{\rm DCB} - \lambda \mathbf{D}\cdot\mathbf{E}_{\rm ext}$, 
where $\mathbf{D}$ is an electric dipole operator.
The operators $\mathbf{T}^{(1)}$ are the solutions of the coupled 
nonlinear equations \cite{ravi-20}
\begin{widetext}
\begin{subequations}
\begin{eqnarray}
	 \langle\Phi^p_a| H_{\rm N}
     + \left [ H_N, {\mathbf T}^{(1)}\right ]
	+ \left [ \left [ H_N, T^{(0)}\right ], {\mathbf T^{(1)}} \right ]
	&+& \frac{1}{2!} \left [ \left [ \left [ H_N, T^{(0)}\right ],
        T^{(0)}\right ], {\mathbf T^{(1)}} \right ] |\Phi_0\rangle  \nonumber \\
	&=&
   \langle\Phi^p_a|\left [ {\mathbf D}, T^{(0)}\right ]
   + \frac{1}{2!}\left [ \left [ {\mathbf D}, T^{(0)}\right ], T^{(0)}
   \right ]|\Phi_0\rangle,  \\
    \langle\Phi^{pq}_{ab}|  H_{\rm N}
   +  \left [ H_N, {\mathbf T^{(1)}} \right ]
   +  \left [ \left [ H_N, T^{(0)}\right ], {\mathbf T^{(1)}} \right]
	&+& \frac{1}{2!}\left [ \left [ \left [ H_N, T^{(0)}\right], T^{(0)}\right ],
      {\mathbf T^{(1)}}\right ]
   + \frac{1}{3!}  \left [ \left [ \left [ \left [ H_N, T^{(0)}\right ],
      T^{(0)}\right ],  T^{(0)}\right ], {\mathbf T^{(1)}}
   \right ]|\Phi_0\rangle  \nonumber \\
	& = &
   \langle\Phi^{pq}_{ab}| \left [ {\mathbf D}, T^{(0)} \right ]
   + \frac{1}{2!} \left [ \left [ {\mathbf D}, T^{(0)}\right ], T^{(0)}
   \right ]|\Phi_0\rangle. 
\end{eqnarray}
  \label{cceq_t1}
\end{subequations}
\end{widetext}
We refer to these equations as the PRCC equations for singles and doubles,
respectively. These equations are linear in ${\mathbf T^{(1)}}$, 
but nonlinear in $T^{(0)}$. More precisely, the left-hand side of 
the singles(doubles) equation contains terms which are two(three) 
orders in $T^{(0)}$. This is to account for the correlation effects 
associated with residual Coulomb interaction more accurately. 
These, as well as unperturbed equation (\ref{ccsd_2v}), are solved 
using the Jacobi method, where to remedy the slow convergence of 
the method we employ direct inversion of the iterated 
subspace (DIIS) \cite{pulay-80}.

The ground state perturbed wavefunction obtained from Eq. (\ref{psi0-ptrb}) 
is then used to calculate the ground state polarizability of Yb and No. 
The dipole polarizability of an atom or ion can be expressed as 
the expectation value of the dipole operator, as
\begin{equation}
  \alpha = -\frac{\langle \widetilde \Psi_0|{\mathbf D}|\widetilde
         \Psi_0\rangle} {\langle \widetilde \Psi_0|\widetilde \Psi_0\rangle}.
\end{equation}
Using Eq. (\ref{psi0-ptrb}), we can write
\begin{equation}
  \alpha = -\frac{\langle \Phi_0|\mathbf{T}^{(1)\dagger}\bar{\mathbf{D}} +
   \bar{\mathbf{D}}\mathbf{T}^{(1)}|\Phi_0\rangle}{\langle\Psi_0|\Psi_0\rangle},
   \label{dbar}
\end{equation}
where $\bar{\mathbf{D}} = e^{{T}^{(0)\dagger}}\mathbf{D} e^{T^{(0)}}$, and 
$\langle \Psi_0| \Psi_0\rangle$ in the denominator is the normalization factor.
Considering the computational complexity, we truncate $\bar{\mathbf{D}}$ as 
well as the normalization factor to second order in $T^{(0)}$. 
From our previous study \cite{mani-10}, using an iterative scheme we 
found that the contribution from the terms with third and higher
orders in $T^{(0)}$ is negligible.

\section{Results and Discussion}

\subsection{Basis set and convergence of properties results}

In order to get accurate results using FSRCC and PRCC theories, it is crucial 
to employ a basis which describes the single-electron wave functions 
and energies accurately. In this work, we use Gaussian type orbitals 
(GTOs) as basis functions \cite{mohanty-91}. The GTO parameters are optimized 
by matching the self-consistent field (SCF) and orbital energies with 
GRASP2K \cite{jonsson-13} and B-spline \cite{oleg-16} results for core-orbitals.
Table \ref{basis} presents the optimized parameters for Yb and No using 
even-tempered basis. Table \ref{ene-no} in the Appendix shows the 
comparison of core-orbitals' energies with B-spline and GRASP2K energies 
for Yb and No.
As evident from the table, for both Yb and No, the energy difference 
between GTO and GRASP2K is less than millihartree. To improve the quality 
of single-particle basis further, we include the corrections from 
the self-energy, through model Lamb-shift operator \cite{shabaev-15}, 
and vacuum polarization, using Uehling potential \cite{uehling-35}.

\begin{table*}
   \caption{The $\alpha_0$ and $\beta$ parameters of the even tempered 
	GTO basis used in our calculations for Yb and No.}
   \label{basis}
   \begin{ruledtabular}
   \begin{tabular}{ccccccccc}
     Atom & \multicolumn{2}{c}{$s$} & \multicolumn{2}{c}{$p$} & 
     \multicolumn{2}{c}{$d$} & \multicolumn{2}{c}{$f$} \\
     \cline{2-3} \cline{4-5} \cline{6-7} \cline{8-9}
     & $\alpha_{0}$  & $\beta$ & $\alpha_{0}$ & $\beta$  
     & $\alpha_{0}$  & $\beta$  & $\alpha_{0}$  & $\beta$ \\
    \hline
     ${\rm Yb}$  &\, 0.00060 &\, 1.9225 &\,0.00415 &\, 1.950 &\, 0.00928 &\, 1.920 &\, 0.00700 &\,1.705 \\
     ${\rm No}$  &\, 0.00750 &\, 1.9980 &\,0.00735 &\, 1.988 &\, 0.00715 &\, 1.955 &\, 0.00650 &\,1.935 \\
   \end{tabular}
   \end{ruledtabular}
\end{table*}

\begin{figure*}
\includegraphics[scale = 0.4, angle=-90]{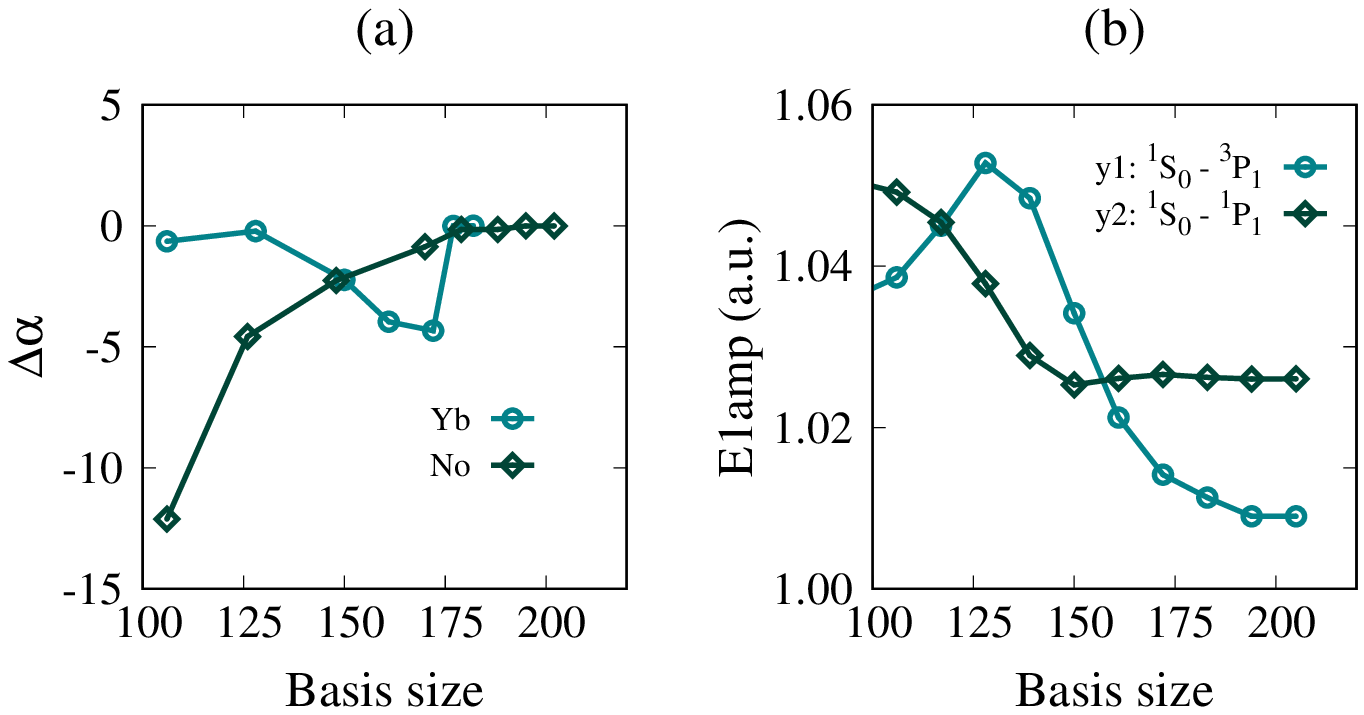}
\caption{Convergence of (a) $\alpha$ of ytterbium and nobelium, (b) E1 transition 
	amplitudes for ${^1}S_{0}\rightarrow {^3}P_{1}$ and ${^1}S_{0}\rightarrow {^1P}_{1}$ 
	transitions of nobelium, and (c) magnetic dipole HFS reduced matrix elements 
	(in the units of $10^{-6}$) of nobelium with basis size.}
\label{fig_conv}
\end{figure*}

Since GTOs form a mathematically incomplete basis \cite{grant-06}, it is 
essential to check the convergence of both unperturbed and perturbed properties
with basis size. 
The convergence trend of $\alpha$, E1 and HFS reduced matrix elements 
with basis size is shown in Fig. \ref{fig_conv}. 
As discernible from the figure, all the properties converge well with 
the basis size. For example, we find that, when the basis is augmented 
from 172 to 177(from 188 to 195) for Yb(No), the change in the value 
of $\alpha$ is $6.1\times10^{-4}$($1.9\times10^{-3}$) a.u. 
Similarly, further augmentation of basis beyond 195 leads to very small 
changes of $1.2\times10^{-3}$ and $1.6\times10^{-4}$ to the transition 
amplitudes of $7s^2\;^{1}S_{0}\rightarrow 7s7p\;^3P_{1}$ 
and $7s^2\;^{1}S_{0}\rightarrow 7s7p\;^1P_{1}$ transitions, respectively, of No.
Therefore, the bases with 188 and 195 orbitals are considered as the 
converged bases for PRCC and FSRCC calculations for No, respectively, and
the corrections from Breit interaction, vacuum polarization and self-energy 
were added to them. 

\subsection{Ionization potential and excitation energy}

In Table \ref{IP}, we present and discuss the ionization potential 
and excitation energy for No. The data from experiments and 
other theoretical calculations are also provided for comparison.
IP and excitation energies are crucial parameters and serve as 
descriptors for the accuracy of the many-body wavefunctions. 
We treated Yb and No as {\em two} valence-electron systems, 
for which IP can be calculated using the difference 
of {\em two}- and {\em one}-electron removal energies, as   
\begin{equation}
	{\rm IP} = E_{ns^2} - E_{ns}.
\end{equation}
Here, $n$ is $6$ and $7$ for Yb and No, respectively. The energies 
$E_{ns^2}$ and $E_{ns}$ are calculated by employing FSRCC theories 
for {\em two}- \cite{mani-11, ravi-21a, palki-24} and {\em one}-valence \cite{mani-17} 
systems, respectively.

\begin{figure}
\includegraphics[height=8.5cm, width=8.5cm, angle=0]{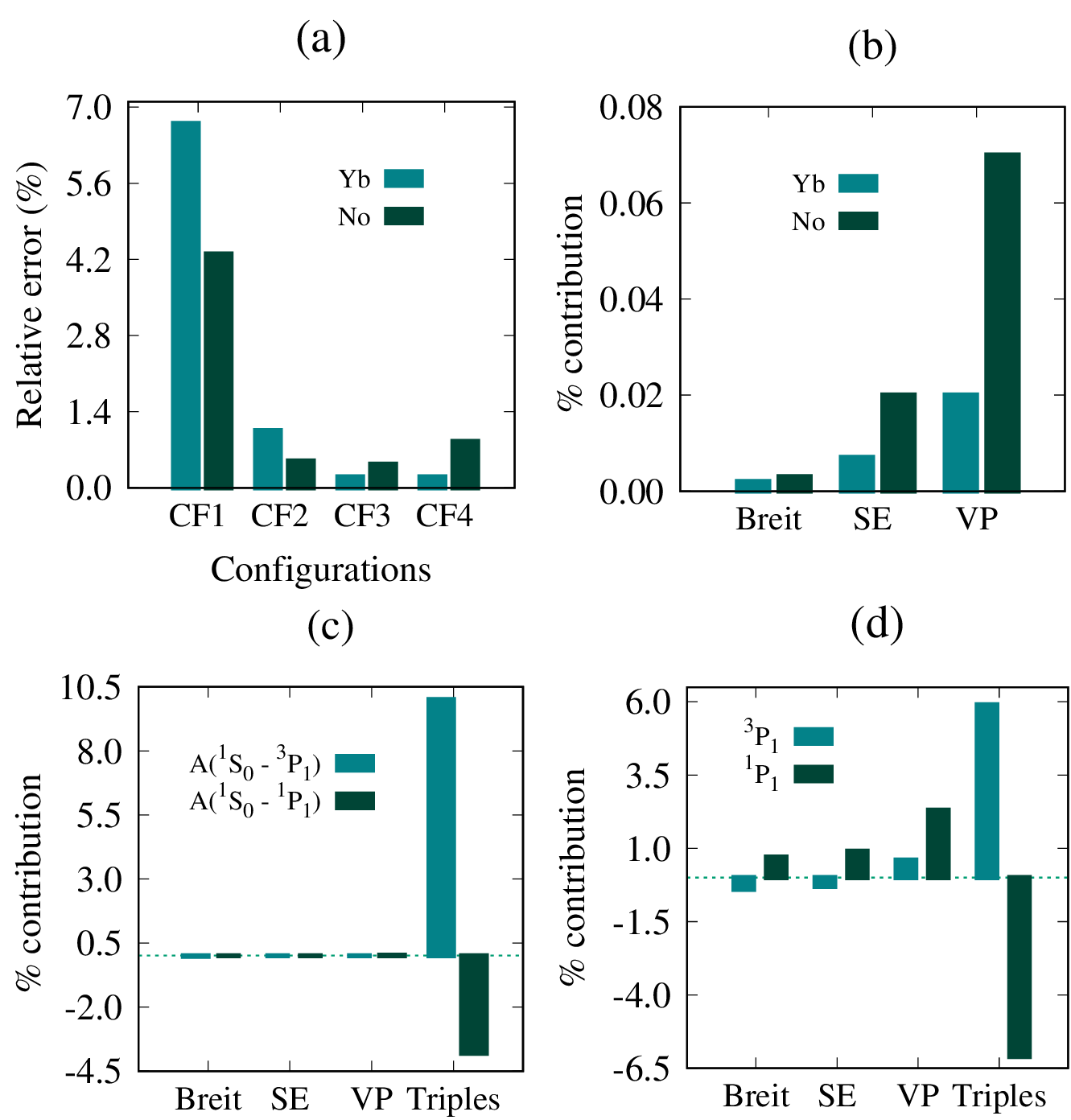}
	\caption{(a) Relative error in the ionization potentials for Yb \Romannum{1}
	and No \Romannum{1}. (b) Contributions from Breit interaction, self-energy and 
	vacuum polarization to ionization potentials of Yb and No. 
	(c), (d) Contributions from Breit interaction, self-energy, vacuum polarization and 
	perturbative triples to transition rates of ${^1}S_{0}\rightarrow{^3}P_{1}$ 
	and ${^1}S_{0}\rightarrow{^1}P_{1}$ transitions and magnetic dipole 
	HFS constants of ${^3}P_{1}$ and ${^1}P_{1}$ states, respectively.} 
\label{fig_unperturbed}
\end{figure}

As evident from the Table \ref{IP-cf}, to account for {\em valence-valence} 
electron correlations more accurately, we also include the higher-energy 
configurations in the model space. For example, for Yb, we start 
with the ground state configuration $6s^2$ and systematically add 
$6s6p$, $6s5d$, and $6s7s$ configurations in the model space. 
As discernible from Fig. \ref{fig_unperturbed}(a), we observed a 
significant improvement in the IP for both Yb and No. The relative error 
has reduced from 6.7 to 0.2\% for Yb, and from 4.3 to 0.8\% for No, 
after including the higher energy configurations. 
This suggests that {\em valence-valence} electron correlation is essential 
to get accurate properties results for multi-valence systems. As can 
be observed from Fig. \ref{fig_unperturbed}(b), the contributions 
from Breit, self-energy and vacuum polarization increase from Yb to No. 
This is expected because relativistic and QED effects are more pronounced
in heavier systems. The combined contribution from Breit and QED effects to IP 
is observed to be $\approx$ 0.03\% and 0.09\% for Yb and No, respectively.

There is a significant variation in the IP values reported from the 
previous calculations for both the systems due to differences in 
the many-body methods employed. For Yb, among all the previous theory 
results, the smallest and largest deviations from the experiment are 
approximately 0.04\% \cite{borschevsky-07} and 18\% \cite{eliav-95}, 
respectively. Among the other works, Refs. \cite{borschevsky-07, cao-01, eliav-95, nayak-06} 
employ a similar methodology as ours. 
Our calculated IP is in good agreement with Refs. \cite{borschevsky-07, nayak-06}. 
The small difference, however, could be attributed to the inclusion 
of higher energy configurations and the corrections from the Breit 
and QED effects in our calculations. Compared to other CCSD calculations \cite{eliav-95, cao-01},
our value is smaller and in better agreement with experiment. 
The remaining results are mainly based on the MCDF calculations and 
show larger deviations from experiment. Our result of IP for Yb is in 
excellent agreement with experiment, with a small relative error 
of 0.2\%. This demonstrates the accuracy of our theory and 
computational framework adopted in the calculations.

Since No and Yb share a similar $(n-2)f^{14}ns^2$ electronic configuration, 
the same electron correlation treatments are also applied to No. Despite the 
competing nature of electron correlations and relativistic effects 
in superheavy elements, our computed IP is in good agreement with the experiment. 
Among all the previous calculations, the result from intermediate Hamiltonian 
based FSRCC calculations \cite{borschevsky-07} is closest to the experiment. 
The reason for this could be attributed to the inclusion of a larger 
model space in Ref. \cite{borschevsky-07}. 
The result from Ref. \cite{cao-03} using CCSD(T) is lower than both
experiment and our calculation. The reason for this could be ascribed to 
the absence of {\em valence-valence} electron correlations due to
few high energy configurations included in the model space. 
The other CC result \cite{dzuba-14} is larger than the experiment and 
ours by $\approx$ 1.8 and $\approx$ 1.0\%, respectively. The reason 
for the difference from our result could be the missing contributions 
from nonlinear CC terms in Ref. \cite{dzuba-14}. The MCDF based 
calculation \cite{liu-07} appears to be more closer to experiment than 
other previous theory calculations except \cite{borschevsky-07}, 
possibly due to an incidental compensation of errors from an 
incomplete treatment of electron correlation.

Beyond IP, we also investigate the transition energies 
for $7s^2{\;^1}S_0 \rightarrow$ $7s7p{\;^1}P_1$ and 
$7s^2{\;^1}S_0 \rightarrow$ $7s7p{\;^3}P_1$ transitions in No.
Experimentally, state ${^1}P_1$ is observed to be located 
at 29961 cm$^{-1}$ \cite{laatiaoui-16} with respect to the ground state, ${^1}S_0$.
The $7s^2{\;^1}S_0\rightarrow7s7p{\;^3}P_1$ transition, however, has 
not been experimentally observed yet, and therefore, theoretical 
calculations become essential in this case.
Our result of 29964 cm$^{-1}$ using $7s^2 + 7s7p$ model configuration
is in excellent agreement with experiment, with a small deviation of 0.01\%. 
However, when an extended model space, $7s^2 + 7s7p + 7s6d + 7s8s$, is 
used we observed a deviation from the experiment. 

Among previous theory results, for ${^1}P_1$ state, the IHFSRCC 
calculation \cite{borschevsky-07} 
is closest to the experiment. Like the case of IP, the MCDF-based 
calculations \cite{liu-07, fritzsche-05} exhibit large variations with 
respect to each other due to model dependencies. 
The result, 30203 cm$^{-1}$, from a combined method of configuration 
interaction and linearized coupled-cluster \cite{dzuba-14} is 
smaller than our result by $\approx$ 1.3\%. 
For ${^3}P_1$ state, our calculation predicts an excitation 
energy of 20630 cm$^{-1}$, which is in excellent agreement with the 
IHFSRCC result of 20454 cm$^{-1}$ \cite{borschevsky-07}. 
Other reported values \cite{dzuba-14, liu-07, fritzsche-05} show 
significant variations due to different treatment of electron 
correlations by many-body methods employed. 
From our calculations, we find combined contribution from 
Breit+QED as $\approx$ 0.5\% and 0.23\% in the excitation
energies of $^3P_1$ and $^1P_1$ states, respectively.

\begin{table*}
\caption{Ionization potential ($\rm cm^{-1}$) of Yb and No, and excitation
    energies ($\rm cm^{-1}$) for $7s^2{\;^1}S_0\rightarrow$ $7s7p{\;^1}P_1$
    and $7s^2{\;^1}S_0\rightarrow$ $7s7p{\;^3}P_1$ transitions in No
    computed using two-valence FSRCC theory. For quantitative analysis
    of electron correlations, contributions from Breit, vacuum polarization
    and self-energy corrections are provided separately.}
  \label{IP}
  \begin{ruledtabular}
  \begin{tabular}{ccccccccc}
      Element/State & {\textrm{FSRCC}} & {\textrm{Breit}} & {\textrm{Vacuum pol.}} & {\textrm{Self-
      energy}}
      & {\textrm{Total}} &  Other theory results  & {\textrm{Expt. \cite{nist}}}   & {\textrm{Error ($\%$)}}\\
      \hline
      \multicolumn{9}{c}{Ionization potential} \\
      Yb \Romannum{1} & $50542$ &  $0.79$ & $ 8.07$ & $ 3.62$ & $50554$ & $49184$$^{a}$,
      $47229$$^{b}$, $41295$$^{c}$,  &  $50443$ & $0.2$ \\
          &          &         &         &          &          & $51109$$^{d}$,
          $48074$$^{e}$, $50463$$^{f}$, &   &       \\
          &          &         &         &          &          & $50552$$^{g}$,
          $48151$$^{h}$, $49684$$^{i}$ &   &      \\

      No \Romannum{1}  & $53900$ &  $1.52$ & $39.27$ & $8.94$  & $53950$   &  $53490$$^{f}$,
          $51055$$^{j}$, $52426$$^{k}$,&  $53443$  & $0.9$   \\
          &            &         &         &          &          & $53701$$^{l}$,
          $53600$$^{m}$, $54390$$^{n}$   &     &    \\

        \hline
        \multicolumn{9}{c}{Excitation energy} \\
        7s7p $^3P_{1}$ & 20630 & 27  & 1 & 45 & 20703  &  $21042$$^{n}$, $20454$
        $^{f}$, $21329$$^{l}$&   & \\
           &&&&&& $20970$$^{o}$ && \\
        7s7p $^1P_{1}$ & 30611 &  13  & 1 & 42 & 30667 & $30203$$^{n}$, $30056$$^{f}$, $30069$$^{l}$
	  & 29961 & 2.4 \\
         &&&&&& $27100$$^{o}$ && \\
        \end{tabular}
        \end{ruledtabular}
        \begin{tabbing}
$^{\rm a}$Ref.\cite{karaccoban-11}[HFR]- Relativistic Hartree-Fock, \\
$^{\rm b}$Ref.\cite{karaccoban-11}[MCHF + BP]- Multiconfiguration Hartree-Fock method
within the framework of the Breit-Pauli Hamiltonian, \\
$^{\rm c}$Ref.\cite{galvez-08}[RNPOEP] - Relativistic numerical parameterized optimized
effective potential method, \\
$^{\rm d}$Ref.\cite{eliav-95}[RFSCC]- Relativistic Fock-space coupled-cluster method, \\
$^{\rm e}$Ref.\cite{migdalek-86}[MC-RHF]- Multiconfiguration relativistic Hartree-Fock, \\
$^{\rm f}$Ref.\cite{borschevsky-07}[IHFCC]- Intermediate-Hamiltonian coupled-cluster method, \\
$^{\rm g}$Ref.\cite{nayak-06}[FSRCC]- Fock-space relativistic coupled-cluster method, \\
$^{\rm h}$Ref.\cite{cao-01}[ACPF + SO]- Ab initio relativistic energy-consistent pseudopotential
multireference averaged coupled-pair functional with spin- \\
orbit corrections, \\
$^{\rm i}$Ref.\cite{cao-01}[CCSD(T)]- Coupled-cluster singles, doubles, and perturbative triples approach, \\
$^{\rm j}$Ref.\cite{cao-03}[ACPF + SO], \\
$^{\rm k}$Ref.\cite{cao-03}[CCSD(T)] \\
$^{\rm l}$Ref.\cite{liu-07}[MCDF] - Multiconfiguration Dirac-Fock, \\
$^{\rm m}$Ref.\cite{sugar-74}[Extrapolation], \\
$^{\rm n}$Ref.\cite{dzuba-14}[CI + all orders] - Configuration interaction method combined with the
linearized single-double coupled-cluster method (all-order), \\
$^{\rm o}$Ref.\cite{fritzsche-05}[MCDF] \\
\end{tabbing}
\end{table*}

\begin{table}
	\caption{Ionization potential (in $\rm cm^{-1}$) for ytterbium and 
    nobelium with increasing model space. To quantitatively assess the 
	{\em valence-valence} electron correlation, cumulative IPs are 
	provided for higher energy configurations in model space in 
	a layer wise manner. }
  \label{IP-cf}
  \begin{ruledtabular}
  \begin{tabular}{lccc}
   Configurations & IP  &   \\
       \hline
       \multicolumn{4}{c}{Yb}\\                 
       \hline  
  CF1 : $6s^2$                &  $47021$  \\   
  CF2 : $6s^2+6s6p$           &  $49914$  \\   
  CF3 : $6s^2+6s6p+6s5d$      &  $50343$  \\   
  CF4 : $6s^2+6s6p+6s5d+6s7s$ &  $50542$  \\   \\ 
       \multicolumn{4}{c}{No}\\                 
       \hline
  CF1 : $7s^2$                &  $51138$  \\   
  CF2 : $7s^2+7s7p$           &  $53183$  \\   
  CF3 : $7s^2+7s7p+7s6d$      &  $53673$  \\   
  CF4 : $7s^2+7s7p+7s6d+7s8s$ &  $53900$  \\
        \end{tabular}
        \end{ruledtabular}
\end{table}

\subsection{Transition rate}

\begin{table*}
\caption{E1 reduced matrix elements (a.u.) and transition rates (s$^{-1}$)
	for $^1S_{0}$ $\rightarrow$ $^3P_{1}$ and $^1S_{0}$ $\rightarrow$ $^1P_{1}$
	transitions in No. For assessment of electron correlations, contributions
	from Breit, QED and perturbative triples are listed separately.
	For comparison, data from experiments and
	other theoretical calculations are also provided.}
\label{transition_rate}
\begin{ruledtabular}
\begin{tabular}{ccccccccc}
\multicolumn{1}{c}{\textrm{States}}  &
\multicolumn{1}{c}{\text{FSRCC}}&
\multicolumn{1}{c}{\text{Breit}}&
\multicolumn{1}{c}{\text{Self-energy}}&
\multicolumn{1}{c}{\text{Vac.-pol.}} &
\multicolumn{1}{c}{\text{Triples}}&
\multicolumn{1}{c}{\text{Total}}&
\multicolumn{1}{c}{\text{Expt.}} &
\multicolumn{1}{c}{\text{Other calculations}}  \\
\hline
\multicolumn{9}{c}{E1 reduced matrix elements} \\
$\langle^3P_1|| D ||^1S_0\rangle$ & -1.0098  & 0.0017   & -0.0001 & -0.0001 & 0.0115 & -0.9968 &        &   \\
$\langle^1P_1|| D ||^1S_0\rangle$ & -3.3734  & -0.0002  &  0.0001 & 0.0053  & 0.0625 & -3.3057 &        &   \\ \\

	\multicolumn{9}{c}{Transition rate ($\times 10^{8}$)} \\
$\langle^3P_1|| D ||^1S_0\rangle$ & 0.0605 & -0.0002 & 0.0001 & 0 & 0.0066 & 0.0670 & & 1.064$^{\rm b}$ \\
$\langle^1P_1|| D ||^1S_0\rangle$ & 2.2045   &   0.0003  &   -0.0001  & -0.0071 & -0.0863
	& 2.1113 & 4.2$^{+2.6}_{-2.8}$$^{\rm a}$ &  3.5$^{\rm c}$, 5.0$^{\rm d}$, 2.7$^{\rm b}$ \\
\end{tabular}
\end{ruledtabular}
\begin{tabbing}
    $^{\rm a}$ Ref.\cite{laatiaoui-16}-Expt.,
    $^{\rm b}$ Ref.\cite{indelicato-07}-MCDF ,
    $^{\rm c}$Ref.\cite{liu-07}-MCDF ,
    $^{\rm d}$Ref.\cite{borschevsky-07}-RCI \\
\end{tabbing}

\end{table*}

In Table \ref{transition_rate}, we present our results on E1 transition 
amplitudes and corresponding transition rates 
for $^1S_{0}$ $\rightarrow$ $^3P_{1}$ and $^1S_{0}$ $\rightarrow$ $^1P_{1}$ 
transitions in No. The transition rate is derived from the reduced 
matrix elements using the relation
\begin{equation}
        A = \frac{2.02613\times10^{18}}{3 \lambda^{3}} S_{\rm E1}, 
\label{trate}
\end{equation}
where $S_{\rm E1} = |\langle {^1}S_0||E1||{^3}P_1 / {^1}P_1\rangle|^2$
is the transition line strength in atomic units computed using FSRCC theory, 
and $\lambda$ is the corresponding wavelength in angstrom.
To quantify different electron correlations, contributions from Breit, QED 
and perturbative triples are provided separately in the table.
Experimentally, the transition rate for $^1S_{0}$ $\rightarrow$ $^1P_{1}$ 
transition was measured for the first time using an atom-at-a-time laser 
resonance ionization spectroscopy \cite{laatiaoui-16}.
However, to the best of our knowledge, there are no experimental results 
on the transition rate for $^1S_{0}$ $\rightarrow$ $^3P_{1}$.
Our result, $2.11\times10^8 s^{-1}$, for $^1S_{0}$ $\rightarrow$ $^1P_{1}$
is within the experimental error bar. 
As evident from the table, previous calculations 
exhibit a large variation. Calculations \cite{liu-07} and \cite{indelicato-07} 
use the same MCDF method; however, the reported transition rates differ 
from each other $\approx$ by 30\%. The reason for this could be the 
inherent limitations associated with the choice of configuration space 
in MCDF calculations. Another theoretical study \cite{borschevsky-07} reports 
a transition rate of 5.0 $\times$ 10$^{8}$ s$^{-1}$ using relativistic 
configuration-interaction (RCI) method, which exceeds our value by 
more than a factor of two. 
This discrepancy could be attributed to the incomplete treatment of 
electron correlation in RCI compared to FSRCC.
For $^1S_{0}$ $\rightarrow$ $^3P_{1}$, to the best of our 
knowledge, there is no experimental data in the literature. From theory 
calculations, however, there is a single result using MCDF 
calculation \cite{indelicato-07}. Our FSRCC transition rate, 
0.07 $\times$ 10$^{8}$ s$^{-1}$, is smaller than MCDF value, 
1.1$\times$ 10$^{8}$ s$^{-1}$ \cite{indelicato-07}.

Fig. \ref{fig_unperturbed}(c) shows the contributions from Breit, 
self-energy, vacuum polarization and perturbative triples 
corrections to the transition rates. We observed a maximum cumulative 
contribution of $\approx$ 0.02\% from Breit and QED to the transition rates.
The contribution from perturbative triples is, however, observed to be
very large. It contributes $\approx$ 10\% and -4\% to the transition 
rates of $^3P_{1}$ and $^1P_{1}$ states, respectively. To the best of 
our knowledge, none of the previous theory calculations on transition 
rates of No incorporate the corrections from triple excitations.

\subsection{Hyperfine splitting and determination of nuclear moments}

To gain an insight into the nuclear structure of odd-mass isotopes of No, 
next we investigate the hyperfine spectra of $^{253}$No (nuclear spin $I = 9/2$). 
It is to be noted that the hyperfine splitting can provide
crucial information on nuclear properties such as nuclear moments, which in 
turn allows the determination of single-particle $g$-factor and nuclear 
deformation. The nuclear moments $\mu$ and $Q$ can be extracted by comparing 
experimentally observed magnetic dipole ($A$) and electric quadrupole 
($B$) HFS constants, respectively, with their theoretical values.

The hyperfine splitting in an atom or ion arises due to the coupling of 
the total electronic angular momentum ($J$) with nuclear spin ($I$). 
The HFS constants $A$ and $B$ in MHz can be expressed 
as \cite{johnsson_book} 
\begin{equation}
	A = \frac{\mu}{I \sqrt{J(J+1)(2J+1)}} \; \langle J||T^{(1)}||J\rangle \times 13074.69,
\end{equation}
and
\begin{equation}
	B = Q \sqrt{\frac{2J(2J-1)}{(2J+1)(2J+2)(2J+3)}} \; \langle J||T^{(2)}||J\rangle \times 469.93,
\end{equation}
respectively. Here, $\mu$ and $Q$ are in the units of nuclear magneton ($\mu_N$) and 
e-barn (eb), respectively. The $T^{(1)}$ and $T^{(2)}$ are rank {\em one} 
and {\em two} irreducible tensor operators, respectively. These can be expressed as
\begin{equation}
	T^{(1)}_q({\mathbf r}) = \frac{-i \sqrt{2}[{\mathbf \alpha \cdot C_{1q}^{(0)}}(\hat r)]}{cr^2} 
	{\rm \;\; and \;\;} 
	T^{(2)}_q({\mathbf r}) = \frac{- C_q^{(2)}(\hat r)}{r^3},
\end{equation}
where $C_{1q}^{(0)}$ is a normalized vector spherical harmonic 
and $C_q^{(2)}$ is a spherical tensor of rank {\em two}. The reduced matrix 
elements $\langle J||T^{(1)}||J\rangle$  and $\langle J||T^{(2)}||J\rangle$
are computed using an all-particle FSRCC theory for two-valence atomic systems, 
developed and demonstrated in our previous work \cite{mani-11}.

\begin{table*}
    \caption{Magnetic dipole and electric quadrupole hyperfine structure
	constants for $^3P_{1}$ and $^1P_{1}$ states of ${^{253}}$No ($I = 9/2$). 
	To get accurate results, corrections from Breit, QED and perturbative 
	triples are also included in the calculations.}
  \label{hfs}
  \begin{ruledtabular}
  \begin{tabular}{lcccc}
   Methods & \multicolumn{2}{c}{$^1P_{1}$} & \multicolumn{2}{c}{$^3P_{1}$}  \\
     \cline{2-3} \cline{4-5}
     &  A (GHz/$\mu_{N}$) & B (GHz/eb) & A (GHz/$\mu_{N}$) & B (GHz/eb)   \\
     \hline
     CCSD                         & -1.467  & 0.961  & 4.505  & -0.752 \\
     CCSD + Breit                 & -1.478  & 0.939  & 4.489  & -0.765  \\
     CCSD + Breit + QED           & -1.524  & 0.897  & 4.505  & -0.794  \\
     CCSD + Breit + QED + Triples & -1.435  & 0.905  & 4.775  & -0.817  \\
     \hline
      &  & Extracted nuclear properties  &  & \\
         & $\mu(\mu_{N})$ & $Q (eb)$ &  & \\
     \hline
     Present work & -0.512 & 3.116 &  &  \\
     Others\cite{raeder-18}-CI + all order   & -0.527 & 5.9   &   &    \\
  \end{tabular}
  \end{ruledtabular}

\end{table*}

In Table \ref{hfs}, we list the values of $A/\mu$ and $B/Q$ from our 
calculations. As evident from the table, our results also incorporate
the corrections from Breit, QED and perturbative triples.
As discernible from Fig. \ref{fig_unperturbed}(d), these interactions 
have significant contributions to HFS constants. For $A$, the largest 
contributions from Breit, self-energy and vacuum polarization are observed
to be $\approx$ 0.7, 0.9 and 2\%, respectively, in the case 
of $^1P_{1}$ state. The largest contribution from perturbative triples
is, however, 6\% for $^1P_{1}$ state. 
Interestingly, for $B$, the contributions from Breit and QED effects 
are observed to be more than the perturbative triples. 
The combined Breit+QED contribution is observed to be about 7\% 
for $B$ of $^1P_{1}$ state, whereas the contribution from the perturbative 
triples is observed to be 0.9\%. The state $^3P_{1}$ is observed to show 
a similar trend for Breit+QED and perturbative triples contributions.

By combining our FSRCC results for $A/\mu$ and $B/Q$ with 
experiment \cite{raeder-18} for $^1P_{1}$, we extract the $\mu$ and $Q$ 
as $-0.512$ $\mu_N$ and $-3.12$ ${\rm eb}$, respectively. 
Our extracted $\mu$ is in good agreement with the CI + all-order
value, -0.527 $\mu_N$, from work \cite{raeder-18}. The reason for
the small difference could, however, be attributed to the inclusion of 
nonlinear CC terms in our method; whereas, CI + all-order \cite{raeder-18} 
is equivalent to linearized coupled-cluster. Our extracted $Q$, however, 
differs by a factor of {\em two} from the CI + all-order 
value, 5.9 ${\rm eb}$ \cite{raeder-18}. 
The observed discrepancy likely arises from a missing factor of half in the 
expression for the quadrupole HFS constant employed 
in Ref.  \cite{raeder-18}. 

\subsection{Isotope shifts and determination of mean square charge radii}

As the isotope shift (IS) is related to the change in the mean square charge 
radius ($\delta\langle {r}^2\rangle$) of the nucleus, one can infer the 
nuclear deformation from the IS measurements.
Considering this, we have computed the isotope
shift parameters for $7s^{2}{\;^1}S_{0}\rightarrow 7s7p{\;^1}P_{1}$
transition in No. It is to be mentioned that this is the only transition
in No for which IS has been measured experimentally \cite{raeder-18}.
To compute IS, we employ the MCDF method as implemented in the GRASP2K \cite{jonsson-13}.
The configuration state functions (CSFs) were generated within 
the framework of MCDF theory \cite{fischer-19} and then frequency 
shifts were calculated using the RIS4 module \cite{ekman-19}.

In Table \ref{iso}, we present our computed mass and field shift parameters. 
As evident from the table, we start with the Dirac-Fock (DF) reference 
configuration and systematically add layer-wise electron correlations by 
incorporating single and double excitations to the active spaces. The first 
model space, referred to as MS1, is defined using the valence reference configurations 
[Rn]5f$^{14}${\;7s$^2$} and [Rn]5f$^{14}${\;7s7p} for even- and odd-parity 
states, respectively. In this case, all core electrons are considered frozen. 
To capture the correlation effects from  the core electrons, we consider a second 
model space, denoted as MS2, in which one of the $5f$-electrons is treated 
as an active electron. Building upon this further, in the next step, we 
consider $6p$ as an active orbital. We refer this model space as MS3.
For all the three model spaces, the correlation layers were systematically 
extended to include the virtual orbitals up to $\{12s,12p,12d,12f,5g\}$ 
for both even and odd parity states. 
As can be expected, the model space MS3, which includes both $5f$ and $6p$ 
electrons as active, yields excitation energy in excellent agreement with 
the experimental value. Considering this, we use MS3 for computing the 
isotope shift parameters.

\begin{figure}
\includegraphics[height=8.0cm, width=8.5cm, angle=0]{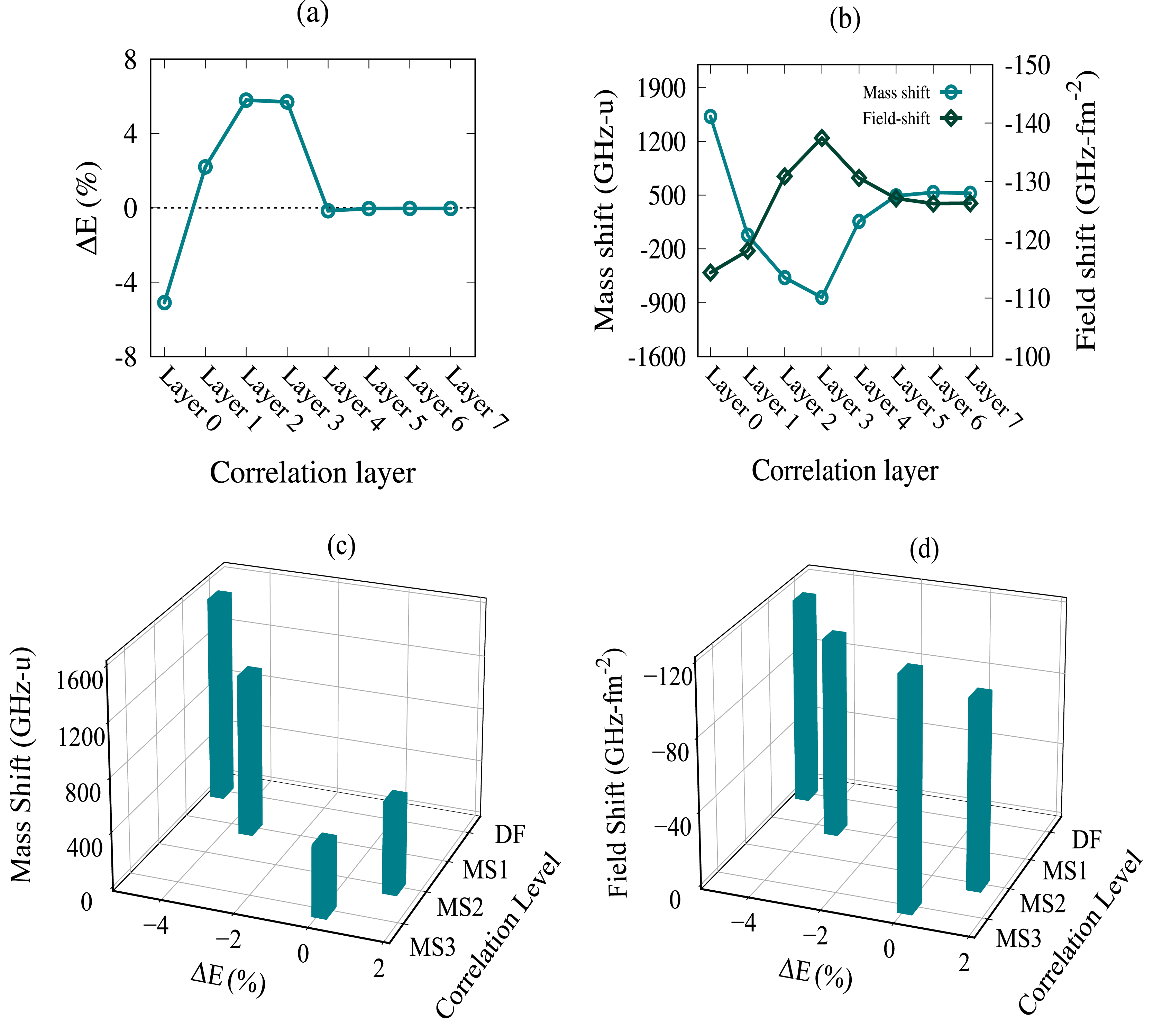}
	\caption{(a), (b) Convergence trend for excitation energy and 
	isotope shift parameters for $7s^2{\;^1}S_0 \rightarrow 7s7p{\;^1}P_1$ transition. 
	(c), (d) The trend of electron correlation with different model configurations.}
\label{fig_iso}
\end{figure}

Figs. \ref{fig_iso}(a) and (b) show the convergence trend for excitation energy 
for ${^1}P_{1}$ and isotope shift parameters for 
$7s^{2}{\;^1}S_{0}\rightarrow 7s7p{\;^1}P_{1}$ transition, respectively.
As discernible from the figures, both the excitation energy and isotope shift
parameters converge well with correlation layer. The converged excitation 
energy is in excellent agreement with the experimental value with a 
small deviation of 0.03\%.  This confirms the accuracy of the many-body 
wavefunctions used in the calculation of isotope shift parameters. 
Figs. \ref{fig_iso} (c) and (d) show the trend of electron correlations 
from different model spaces to mass and field-shift parameters, respectively. 
As can be observed from the figures, there is  a large contribution from 
the $5f$ core electrons to the mass shift ($M_{\rm s}$) parameter. It reduces
the DF value by $\approx$ 55\%. As can be expected, the preceding core, 
$6p$, has a less contribution than $5f$, and reduces the mass shift parameter 
further by 22\%. The field shift ($F_{\rm s}$) parameter also show a 
trend of opposite contribution from $5f$, however, with much lesser 
magnitude. Unlike $M_{\rm s}$, for $F_{\rm s}$, $6p$ core electrons 
have contribution in the opposite phase to $5f$, and hence increases the 
value further.
From other theory calculations, we found only one reported value 
of $M_{\rm s}$, using the MCDF method \cite{raeder-18}. The reported value, 
$-1044\pm400$ \cite{raeder-18}, has a large error of $\approx$ 38\%. 
Our computed value, $524.4$, is almost half of the calculation \cite{raeder-18}, 
and has an opposite sign. Ref. \cite{raeder-18} also reports the
value of $F_{\rm s}$ using different methods. All the reported values, 
however, have large errors. Among all the methods, the MCDF result is 
the largest. Our computed value, $-126.2$, is more closer to the 
MCDF result \cite{raeder-18}.

\begin{table}
    \caption{Transition energy ($\Delta E$), mass shift ($M_{\rm s}$)
    and field shift ($F_{\rm s}$) constants
    for $7s^2{\;^1}S_0 \rightarrow 7s7p{\;^1}P_1$ transition in
    nobelium calculated using MCDF method. Results from the layer-wise
    augmentation of configuration space are provided to
    assess the nature of electron correlations.}
\label{iso}
\begin{ruledtabular}
\begin{tabular}{cccccc}
    Layer & $\Delta$E (cm$^{-1}$) & $M_{\rm s}$(GHz u)
        &  $F_{\rm s}$ (GHz/fm$^{2}$)\\
    \hline
    0$^{a}$  & 28437  & 1525.45 &  -114.37    \\
        1$^{b}$  & 30623  &  -21.28 &  -118.17     \\
        2$^{c}$  & 31692  &  -573.38 &  -130.87      \\
        3$^{d}$  & 31654  &  -832.15 &  -137.43       \\
        4$^{e}$  & 29912  &   158.60 &  -130.59        \\
        5$^{f}$  & 29949  &   491.12 &  -127.07         \\
        6$^{g}$  & 29953  &   535.10 &  -126.20          \\
        7$^{h}$  & 29953  &   524.40 &  -126.24         \\
        \hline
    Other results     &         & -1044(400)$^{i}$
    & -95.8(7)$^{j}$, -104(10)$^{k}$, -94(25)$^{l}$  \\
    &&& -99(15)$^{m}$, -113(25)$^{n}$ \\
\end{tabular}
\end{ruledtabular}
\footnotetext{Layer 0 - DF}
\footnotetext{Layer 1 - Even: \{8s,7p,6d,6f\}, Odd: \{8s,8p,6d,6f\}}
\footnotetext{Layer 2 - Even: \{9s,8p,7d,7f,5g\}, Odd: \{9s,9p,7d,7f\}}
\footnotetext{Layer 3 - Even: \{10s,9p,8d,8f,5g\}, Odd: \{10s,10p,8d,8f\}}
\footnotetext{Layer 4 - Even: \{11s,10p,9d,9f,5g\}, Odd: \{11s,11p,9d,9f,5g\}}
\footnotetext{Layer 5 - Even: \{12s,11p,10d,10f,5g\}, Odd: \{12s,12p,10d,10f,5g\}}
\footnotetext{Layer 6 - Even: \{12s,12p,11d,11f,5g\}, Odd: \{12s,12p,11d,11f,5g\}}
\footnotetext{Layer 7 - Even: \{12s,12p,12d,12f,5g\}, Odd: \{12s,12p,12d,12f,5g\}}
\footnotetext{Ref.\cite{raeder-18} - MCDF}
\footnotetext{Ref.\cite{raeder-18} - CI + all orders}
\footnotetext{Ref.\cite{raeder-18} - CI + MBPT}
\footnotetext{Ref.\cite{raeder-18} - CIPT}
\footnotetext{Ref.\cite{raeder-18} - FSCC}
\footnotetext{Ref.\cite{raeder-18} - MCDF}
\end{table}

Combining our computed $M_{\rm s }$ and $F_{\rm s }$ parameters with 
experimental isotope shift for $7s^2{\;^1}S_0 \rightarrow 7s7p{\;^1}P_1$
transition, we extracted the change in the mean square charge
radii of $^{252, 253, 255}$No nuclei relative to
$^{254}$No, using the relation \cite{raeder-18}
\begin{equation}
	\delta {\nu}_{\rm IS}^{\rm AA'} = M_{\rm s } \frac{(A' - A)}{AA'}
	+ F_{s} \delta \langle r^{2} \rangle^{AA'}.
\end{equation}
Here, $\delta {\nu}_{\rm IS}^{\rm AA'}$ is the total isotope shift of
an isotope with mass number $A^{'}$ compared to an isotope with mass
number $A$. Using this expression, we extracted the change in the
mean square charge radii for $^{252, 253, 255}$No nuclei relative
to $^{254}$No as $-0.080$, $-0.0535$ and $-0.0535$ fm$^{2}$, respectively.
Our obtained values are smaller than the values, $-0.105$ \cite{raeder-18}, 
$-0.075$ \cite{raeder-18} and $-0.080$ \cite{warbinek-24} fm$^{2}$, respectively. 
The reason for this could be attributed to our slightly larger value 
of $F_{\rm s}$. It is to be noted that, Ref. \cite{raeder-18} has used 
the CI+all-order value, $-95.8$, in the extraction, which is smaller 
than our value.

\subsection{Electric dipole polarizability}

\begin{table}
	\caption{The value of $\alpha$ (in a. u.) from PRCC calculation compared 
	with other theoretical data in the literature.}
  \label{final-alpha}
  \begin{ruledtabular}
  \begin{tabular}{clcr}
      Element & \multicolumn{2}{c}{\textrm{Present work}} & {\textrm{Other cal.}}\\
            \hline
           & {\textrm{Method}}& $\alpha$ &  \\
            \hline
      Yb   & DF                & $122.911$  &  $144.6\pm5.6^{\rm a}$,     \\
           & PRCC              & $145.397$  &  $140.7\pm7.0^{\rm b}$, $142.6^{\rm d}$, \\
           & PRCC(T)           & $142.814$  &  $141\pm6^{\rm c}$, $138.9^{\rm e}$,  \\
           & PRCC(T)+Breit     & $141.952$  &  $142^{\rm f}$,$144^{\rm g}$,$141\pm2^{\rm h}$,    \\
           & PRCC(T)+Breit+QED & $142.154$  &  $141\pm4^{\rm i}$, \\ 
           & Estimated         & $142.160$  &  $135.73^{\rm k}$, $152.9^{\rm l}$, $143^{\rm m}$, \\
	   & Recommended            & $142.2$   &  $157.3^{\rm n}$, $151.0^{\rm o}$, $136\pm5^{\rm p}$,   \\ 
           &                   & $\pm4.1$  &  $147\pm20^{\rm q}$, $139.3\pm5.9^{\rm r}$  \\
            \hline
     No    & DF                & $104.288$  &  $110.8\pm5.5^{\rm b}$, $105.4^{\rm e}$,  \\
           & PRCC              & $107.119$  & $114^{\rm f}$, $107.77^{\rm k}$, ,  \\
	       & PRCC(T)           & $109.171$  & $110\pm6^{\rm s}$, $115.6^{\rm t}$ \\
           & PRCC(T)+Breit     & $108.660$  &  \\
	       & PRCC(T)+Breit+QED & $108.891$  &  \\
	       & Estimated         & $108.715$  &  \\
	       & Recommended       & $108.7$      &  \\ 
	       &                   & $\pm3.2$       &  \\ 
        \end{tabular}
        \end{ruledtabular}
\begin{tabbing}
  $^{\rm a}$Ref.\cite{sahoo-08}[CCSD], \ 
  $^{\rm b}$Ref.\cite{peter-09}[CCSD(T)],  \\
  $^{\rm c}$Ref.\cite{dzuba-10}[CI+MBPT], \  
  $^{\rm d}$Ref.\cite{buchachenko-11}[CCSD(T)], \\  
  $^{\rm e}$Ref.\cite{dzuba-14a}[CI+MBPT+RPA], \\ 
  $^{\rm f}$Ref.\cite{dzuba-16}[R-RPA: Relativistic random phase approximation], \\
  $^{\rm g}$Ref.\cite{zhang-09}[R-CCSD], \ 
  $^{\rm h}$Ref.\cite{safronova-12}[CI+MBPT+RPA], \\ 
  $^{\rm i}$Ref.\cite{wang-98}[DHF+Breit+QED],  \  
  $^{\rm k}$Ref.\cite{yoshizawa-16}[DFT],   \
  $^{\rm l}$Ref.\cite{buchachenko-06}[CCSD(T)], \\  
  $^{\rm m}$Ref.\cite{zhang-07}[CCSD(T)],  \ 
  $^{\rm n}$Ref.\cite{xchu-07}[DFT: Density functional theory], \\  
  $^{\rm o}$Ref.\cite{buchachenko-07}[AQCC: Averaged quadratic coupled cluster],   \\ 
  $^{\rm p}$Ref.\cite{sahoo-18}[CCSD(T)],  \     
  $^{\rm q}$Ref.\cite{lei-15}[Exp.], \
  $^{\rm r}$Ref.\cite{beloy-12}[Exp.],  \\      
  $^{\rm s}$Ref.\cite{dzuba-14}[CI+all order],  \\ 
  $^{\rm t}$Ref.\cite{martin-16}[DFT-DKH: Density functional theory solved using \\
  Douglas-Kross-Hess Hamiltonian],  \\       
\end{tabbing}
\end{table}

\begin{table}[ht]
    \caption{Contributions to $\alpha$ (in a.u.) from different terms 
	in the PRCC theory.}
    \label{termwise}
    \begin{center}
    \begin{ruledtabular}
    \begin{tabular}{lrr}
        Terms + H.c. & \multicolumn{1}{r}{$\rm{Yb}$}
        & \multicolumn{1}{r}{$\rm{No}$}  \\
        \hline
        $\mathbf{T}_1^{(1)\dagger}\mathbf{D} $            & $186.3212$   & $146.6744$  \\
        $\mathbf{T}_1{^{(1)\dagger}}\mathbf{D}T_2^{(0)} $ & $ -9.7606$   & $ -7.5187$  \\
        $\mathbf{T}_2{^{(1)\dagger}}\mathbf{D}T_2^{(0)} $ & $ 12.6231$   & $  8.3248$  \\
        $\mathbf{T}_1{^{(1)\dagger}}\mathbf{D}T_1^{(0)} $ & $-14.6273$   & $-13.1906$  \\
        $\mathbf{T}_2{^{(1)\dagger}}\mathbf{D}T_1^{(0)} $ & $  1.4426$   & $  1.2839$  \\
        Normalization                                     & $ 1.21047$   & $ 1.26563$  \\
        Total                                             & $145.3973$   & $107.1196$  \\
    \end{tabular}
    \end{ruledtabular}
    \end{center}
\end{table}

In Table \ref{final-alpha}, we have provided the final value of $\alpha$ 
for the ground state, $^1S_0$, of Yb and No computed using PRCC theory. 
To understand the trend of electron correlations embedded in PRCC theory, 
we have provided separate contributions at different levels of the theory. 
DF represents the Dirac-Fock contribution and, as to be expected, has
the dominant contribution. 
The contribution is calculated by replacing $\mathbf{T}^{(1)}$ and 
$\bar{\mathbf{D}}$ in Eq. (\ref{dbar}) with bare dipole operators.
For both the atoms, the DF values are smaller than the final $\alpha$.
We observe DF contributions of $\approx$ 88\% and 96\% of the total 
value for Yb and No, respectively.
The PRCC refers to the contribution from perturbed relativistic coupled-cluster 
theory where residual Coulomb interaction is accounted to all orders and 
the effect of external electric field is considered up to the first-order 
of perturbation. 
The PRCC(T) includes the contribution from perturbative triples. 
The PRCC(T)+Breit+QED includes the contributions from Breit and 
QED corrections along with perturbative triples.
And the term {\em Estimated} refers to the estimated cumulative 
contribution from $i$, $j$ and $k$ symmetry orbitals.

For Yb, our recommended value of $\alpha$ is within the experimental 
uncertainty \cite{lei-15, beloy-12}. In terms of other theory 
calculations, $\alpha$ for ground state of Yb is calculated using 
various methods such as relativistic 
coupled-cluster (RCC) \cite{sahoo-08,peter-09,buchachenko-11,zhang-09,
buchachenko-06,buchachenko-06, zhang-07,sahoo-18}, 
CI+MBPT \cite{dzuba-10,dzuba-14a,safronova-12}, CI+all-order \cite{dzuba-14}, 
RPA \cite{dzuba-16} and DFT \cite{yoshizawa-16,xchu-07, martin-16}.    
However, there is a large variation in the $\alpha$ values reported using RCC 
theories and also across other methods. For example, the value reported 
in Ref. \cite{sahoo-18} is $\approx$ $12$\% smaller 
than Ref. \cite{buchachenko-06}, while both of these works have 
used CCSD(T) method. Our recommended 
value 142.2$\pm$4.1 is consistent with most of the RCC based
calculations. Our result is also consistent with CI+MBPT and 
based calculations \cite{dzuba-10,dzuba-14a,safronova-12}.

For No, to the best of our knowledge, there is no experimental data
for ground state $\alpha$. However, we could find six previous calculations 
for comparison. Out of these, 
Ref. \cite{peter-09} uses a CCSD(T) method, similar to ours, however, with 
a difference that we also include the corrections from the QED effects in our 
calculations. Our recommended value 108.7$\pm$3.2 is consistent 
with the value, 110.8$\pm$5.5, in Ref. \cite{peter-09}. As other important results 
for ground state $\alpha$ of No, Dzuba {\em et al.} has reported the values
using RHF+RPA \cite{dzuba-16}, CI+MBPT+RPA \cite{dzuba-14a} and 
CI+all-order \cite{dzuba-14} methods. Our recommended value is closer 
to the CI+all-order \cite{dzuba-14} value, $110$. The reason for this 
could be attributed to the more accurate treatment of electron 
correlations in CI+all-order than other two methods.
The remaining two calculations \cite{yoshizawa-16} and \cite{martin-16}
are using the density functional theory based calculations, however,
differ from each other by $\approx$ 7\%. 


\subsubsection{Electron correlations embedded in PRCC}

To analyze the electron correlation effects embedded in PRCC in more detail, 
we have separated the contribution into five different terms and listed them in
Table \ref{termwise}. For both the atoms, the most dominant contribution is 
from the leading order (LO) term $[\mathbf{T}_1^{(1)\dagger}\mathbf{D} + {\rm H.c.}]$. 
It is as expected because this term subsumes the contributions from DF and 
dominant RPA effects.
Its contribution is $\approx 28.1$\% and $\approx 36.9$\% more than the total 
$\alpha$ for Yb and No, respectively. The next leading order (NLO) contribution 
is observed from the term $[\mathbf{T}_1{^{(1)\dagger}}\mathbf{D}T_1^{(0)}+{\rm H.c.}]$. 
In contrast to the LO term, the contribution is opposite in phase with $\approx -10.1$\% 
and $\approx -12.3$\% of total $\alpha$ for Yb and No, respectively.
Next to NLO term is $[\mathbf{T}_2{^{(1)\dagger}}\mathbf{D}T_2^{(0)} + {\rm H.c.}]$, 
and it contributes $\approx 8.6$\% and $\approx 7.8$\%, respectively for Yb and No.
The term $[\mathbf{T}_1{^{(1)\dagger}}\mathbf{D}T_2^{(0)} + {\rm H.c.}]$ 
also has a significant contribution of $\approx -7$\% for each atom.
The remaining terms collectively contribute $\approx 1$\% for both 
the atoms.

\begin{figure*}
\includegraphics[scale=0.35, angle=-90]{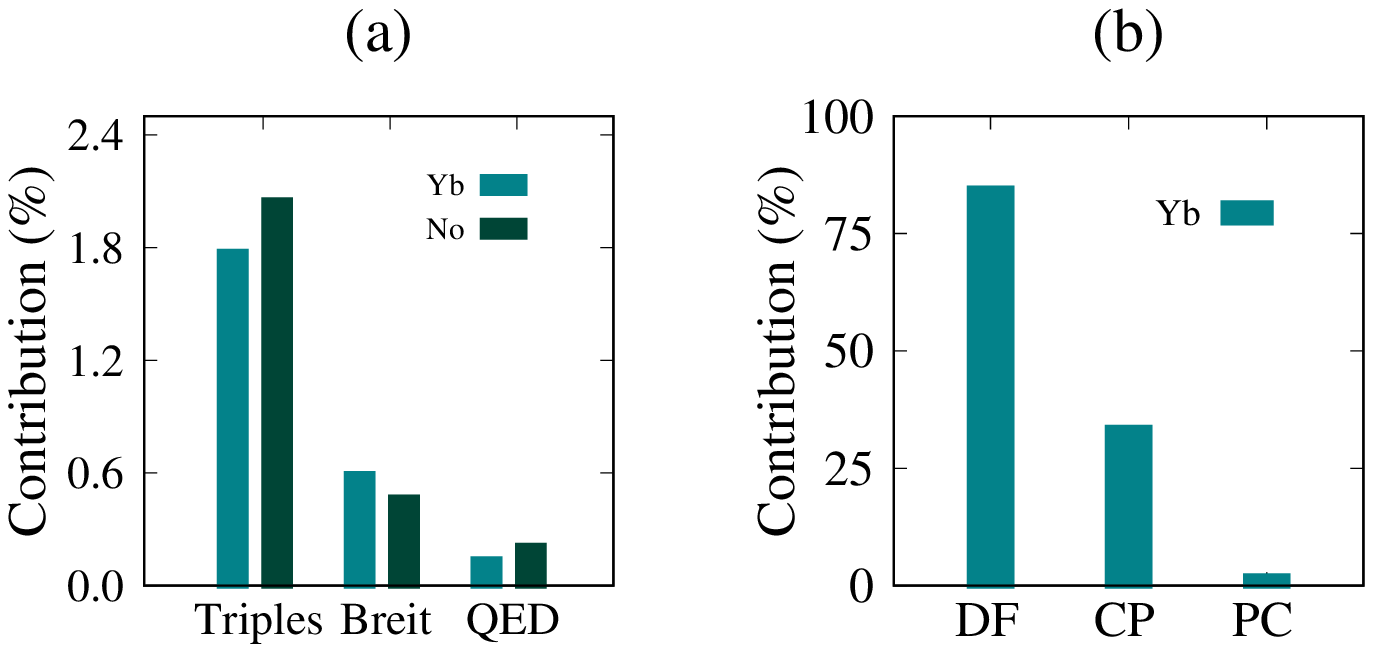}
 \caption{(a) Contributions from Breit, QED and perturbative triples to the 
	ground state $\alpha$ of Yb and No. (b), (c) The percentage contributions 
	from DF, CP, and PC to the $\alpha$ of Yb and No.}
\label{fig_CP-PC}
\end{figure*}

To get further insight into the electron correlation, next we examine the 
contributions from core-polarization (CP) and pair-correlation (PC) effects. 
To extract the CP contribution, we used the LO term 
$[\mathbf{T}_1^{(1)\dagger}\mathbf{D} + {\rm H.c.}]$, which subsumes 
the dominant CP contribution. Some CP effects are also included in
the NLO term $[\mathbf{T}_1{^{(1)\dagger}}\mathbf{D}T_1^{(0)} + {\rm H.c.}]$. 
To estimate the pair-correlation (PC) effect, we consider the combined 
contributions from the terms $[\mathbf{T}_1{^{(1)\dagger}}\mathbf{D}T_2^{(0)} + {\rm H.c.}]$ 
and $[\mathbf{T}_2{^{(1)\dagger}}\mathbf{D}T_2^{(0)} + {\rm H.c.}]$.
The percentage contributions from DF, CP and PC are shown in Fig. \ref{fig_CP-PC} 
for both the atoms.
As can be expected, DF has the most dominant contributions of 84.5\% and 
97.1\% of the total $\alpha$, respectively, for Yb and No. The CP contributes 
$\approx$ 33.5\% and 28.2\%, respectively, for Yb and No, whereas the 
contributions from PC are $\approx$ 2\% and 0.8\%, respectively. The reason
for the smaller contribution from PC is the cancellation due to opposite 
contributions from $[\mathbf{T}_1{^{(1)\dagger}}\mathbf{D}T_2^{(0)} + {\rm H.c.}]$
and $[\mathbf{T}_2{^{(1)\dagger}}\mathbf{D}T_2^{(0)} + {\rm H.c.}]$ terms.

\begin{figure}
\includegraphics[height=8.8cm, angle=-90]{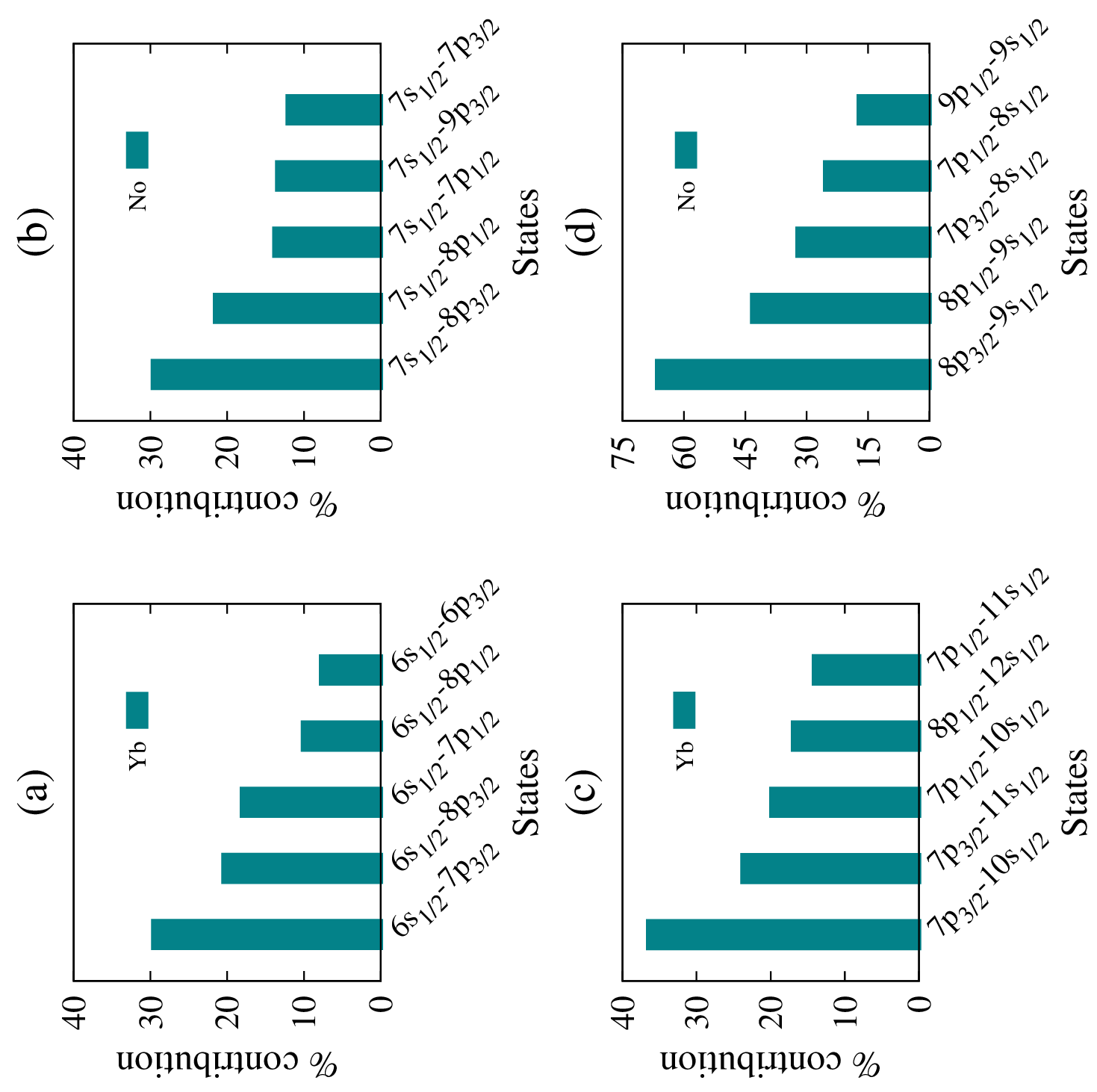}
	\caption{(a), (b) Five largest percentage contributions from 
	the dipolar mixing of core and virtuals extracted from LO 
	term for Yb and No. 
	(c), (d) Five largest percentage contribution from the dipolar 
	mixing of virtual-virtual orbitals of NLO terms for Yb and No.} 
\label{fig_orb_cont}
\end{figure}

Next, to get further insight into the correlation from individual orbitals,
we identified core and virtual orbitals which have dominant contributions. 
Fig. \ref{fig_orb_cont} shows the five largest dipolar mixings between 
{\em core-virtual} and {\em virtual-virtual} pairs, extracted from 
the LO and NLO terms, respectively. As discernible from the panels (a) and 
(b) of the figure, as can be expected, $\approx$ 86\% and 91\% of
contributions, respectively, for Yb and No come from the outermost 
orbitals $6s_{1/2}$ and $7s_{1/2}$. For Yb, $6s_{1/2}$ contributes 
through dipolar mixing with $7p_{3/2}$, $8p_{3/2}$, $7p_{1/2}$,
$8p_{1/2}$ and $6p_{3/2}$, whereas for No, it comes through the mixing
with $8p_{3/2}$, $8p_{1/2}$, $7p_{1/2}$, {\bf $9p_{3/2}$} 
and $7p_{3/2}$ orbitals. As the dominant contributions 
from {\em virtual-virtual} pairs in NLO term, for Yb, $\approx 94$\% 
contribution is from the mixing of $7p$ with $10s_{1/2}$ and $11s_{1/2}$ 
orbitals (panel (c)). The second largest contribution of $\approx 14$\% 
is from the mixing between $8p_{1/2}$ and $12s_{1/2}$ virtuals.
Similarly, for No (panel (d)), $\approx$ 127\% of NLO contribution 
comes from the dipolar mixing of $9s$ with $8p$ and $9p$ orbitals.
As the second largest contribution, we observed a contribution 
of $\approx 57$\% from the mixing of $8s$ with $7p$ states.

\begin{figure}
\includegraphics[height=8.8cm, angle=-90]{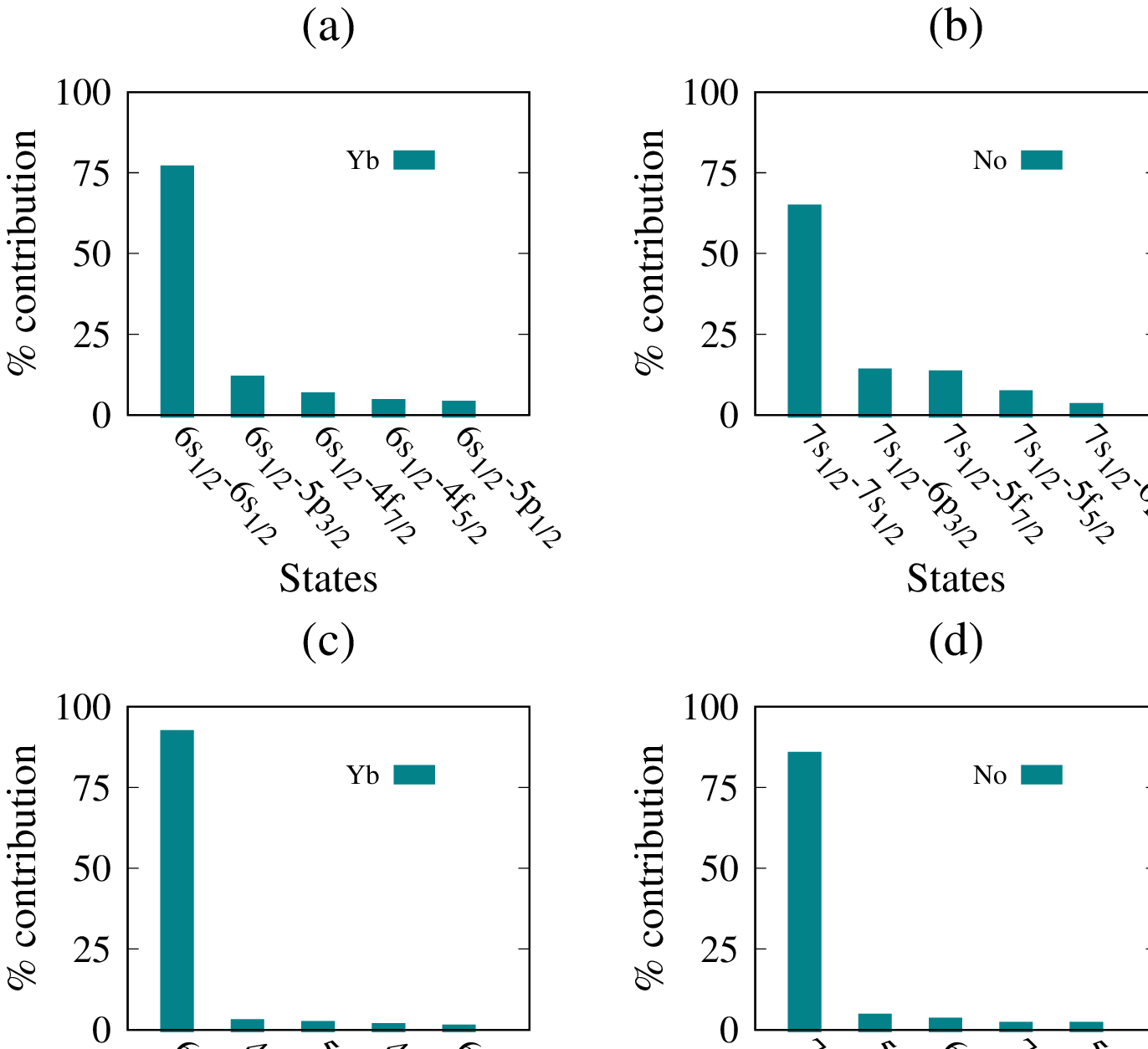}
	\caption{Five largest percentage contribution from the
	{\em core-core} pairs in the terms ${\bf T_1}^{(1)\dagger}DT_2^{(0)}$+ H.c. 
	(panels (a) and (b)) and ${\bf T_2}^{(1)\dagger}DT_2^{(0)}$+ H.c. 
	(panels (c) and (d)) for Yb and No.} 
\label{fig_cc}
\end{figure}

Table \ref{tab_cc} shows the five leading order {\em core-core} pair 
contributions from the terms $[{\bf T_1^{(1)\dagger}} D \ T_2^{(0)} + {\rm H.c.}]$ 
and $[{\bf T_2^{(1)\dagger}}D \ T_2^{(0)} + {\rm H.c.}]$. The percentage 
contribution from the same is shown in Fig. \ref{fig_cc} for an easy 
assessment. As discernible from the panels (a) and (b) of the figure, from 
the term $[{\bf T_1^{(1)\dagger}}DT_2^{(0)}$+ H.c.], the most dominant contribution 
of $\approx$ 76\%(64\%) is from the $6s_{1/2}-6s_{1/2}$($7s_{1/2}-7s_{1/2}$) 
core pairs for Yb (No). The remaining contribution of 24\%(36\%) comes from the 
pair of $6s_{1/2}$($7s_{1/2}$) with $5p_{3/2}$, $4f_{7/2}$, $4f_{5/2}$, 
and $5p_{1/2}$($6p_{3/2}$, $5f_{7/2}$, $5f_{5/2}$, and $6p_{1/2}$) cores for Yb(No). 
The term $[{\bf T_2^{(1)\dagger}}DT_2^{(0)}$+H.c.] also shows a similar trend 
where the dominant contributing {\em core-core} pairs are $6s_{1/2}-6s_{1/2}$ 
and $7s_{1/2}-7s_{1/2}$ for Yb and No, respectively, and they contribute 
$\approx$ 92\% and 85\% for Yb and No, respectively (panels (c), (d)). 
Among the remaining cores, $5p_{3/2}$, $4f_{7/2}$, and $4f_{5/2}$($6p_{3/2}$, 
$5f_{7/2}$, and $5f_{5/2}$) with $6s_{1/2}$($7s_{1/2}$) core pairs 
contribute $\approx$ 6\% and 10\% to $\alpha$ for Yb(No).

\begin{table}
	\caption{Five leading {\em core-core} contributions (in a.u.)
	corresponding to the pair-correlation terms 
	${\bf T_1^{(1)\dagger}} D \ T_2^{(0)} + {\rm H.c.}$ and 
	${\bf T_2^{(1)\dagger}}D \ T_2^{(0)} + {\rm H.c.}$. }
\begin{center}
\begin{ruledtabular}
\begin{tabular}{ccc}
    $\rm{Yb}$  &&  $\rm{No}$ \\
 &   ${\bf T_1^{(1)\dagger}} D \ T_2^{(0)} + {\rm H.c.} $ \\
                       \hline
    $-7.420\ \ (6s_{1/2},6s_{1/2})$  &&  $-4.868\ \ (7s_{1/2},7s_{1/2})$ \\
    $-1.102\ \ (6s_{1/2},5p_{3/2})$  &&  $-1.024\ \ (7s_{1/2},6p_{3/2})$ \\
    $-0.604\ \ (6s_{1/2},4f_{7/2})$  &&  $-0.980\ \ (7s_{1/2},5f_{7/2})$ \\
    $-0.398\ \ (6s_{1/2},4f_{5/2})$  &&  $-0.520\ \ (7s_{1/2},5f_{5/2})$ \\
    $-0.352\ \ (6s_{1/2},5p_{1/2})$  &&  $-0.214\ \ (7s_{1/2},6p_{1/2})$ \\ 
                                \\
 &   ${\bf T_2^{(1)\dagger}}D \ T_2^{(0)} + {\rm H.c.} $ \\
                       \hline
    $11.490\ \ (6s_{1/2},6s_{1/2})$  &&  $7.216\ \ (7s_{1/2},7s_{1/2})$ \\
    $ 0.306\ \ (4f_{7/2},6s_{1/2})$  &&  $0.346\ \ (5f_{7/2},7s_{1/2})$ \\
    $ 0.230\ \ (5p_{3/2},6s_{1/2})$  &&  $0.242\ \ (6p_{3/2},7s_{1/2})$ \\
    $ 0.148\ \ (4f_{5/2},6s_{1/2})$  &&  $0.138\ \ (7s_{1/2},6p_{3/2})$ \\
    $ 0.092\ \ (6s_{1/2},5p_{3/2})$  &&  $0.134\ \ (5f_{5/2},7s_{1/2})$ \\ 
\end{tabular}
\end{ruledtabular}
\end{center}
\label{tab_cc}
\end{table}

\subsubsection{Corrections from Breit, QED and perturbative triples}

Fig. \ref{fig_CP-PC}(a) shows the percentage contributions from Breit, 
QED and perturbative triples to $\alpha$. 
As discernible from the figure, the Breit contribution for No is
smaller than Yb. A similar trend was also observed in 
the case of group-13 ions \cite{ravi-20} where higher $Z$ atoms were 
observed to have smaller Breit contributions. However, consistent with our
previous studies on group-13 ions \cite{ravi-20} and superheavy 
elements \cite{ravi-21b}, the QED correction is larger in No than Yb.
In terms of percentage, Breit and QED contribute $\approx$ 0.47(0.59)\% 
and 0.21(0.14)\%, respectively, to $\alpha$ for No(Yb). As can be observed 
from the figure, perturbative triples have significant contributions. They 
contribute  $\approx$ $-1.8$\% and 2.2\% for Yb and No, respectively.
From previous calculations, for Yb, we find a mixed trend of contributions 
from  perturbative triples, $-4.51$\% \cite{peter-09} and $-3.89$\% \cite{zhang-07}, 
$0.57$\% \cite{sahoo-18}. Our result is consistent in terms of 
sign with Refs. \cite{peter-09, zhang-07}, however, smaller in magnitude.

\section{Theoretical uncertainty} 

The theoretical uncertainty in our computed transition rates, Eq. (\ref{trate}), 
depends on the uncertainties in E1 reduced matrix elements and the 
wavelengths associated with transitions.
Whereas for HFS constants, it depends only on the uncertainties in the 
HFS matrix elements. For this, have identified five different sources 
which can contribute to the uncertainty of E1 and HFS reduced matrix elements. 
The first source of uncertainty is due to the truncation of the basis
in our calculations. As discussed in the basis convergence 
section, the change in the E1 reduced matrix elements is of the order 
of $10^{-3}$ or smaller with basis size. Since this is a very small 
change, we may neglect this uncertainty. The second source of 
uncertainty is from the truncation of the dressed operator at 
the second order of $T^{(0)}$ in the properties calculation \cite{mani-11}.
In our earlier work \cite{mani-10}, using an iterative scheme, we have 
shown that the terms with third and higher orders in $T^{(0)}$ contribute 
less than 0.1\%. So, we consider 0.1\% as an upper bound for this source. 
The third source of uncertainty is due to the partial inclusion of 
triple excitations in the properties calculation. Since the perturbative 
triples account for the leading order terms of triple excitation, 
the contribution from remaining terms will be small. Based on the 
analysis from our previous works \cite{ravi-20, chattopadhyay-15}, 
we estimate the upper bound from this source as 0.72\%. 
The fourth source of uncertainty 
could be associated with the frequency-dependent 
Breit interaction which is not included in the present
calculations. However, in our previous work \cite{chattopadhyay-14}, using a
series of computations using GRASP2K we estimated an upper bound on this
uncertainty as $0.13$\% in Ra. So, for the present work, we take $0.13$\% as
an upper bound from this source. 
The fifth source of uncertainty arises due to the use of incomplete model 
space in our calculations to avoid the intruder states. Based on the analysis 
of the model dependent contributions, we estimate an upper bound to 
this source of uncertainty as 0.5\%.
There could be other sources of theoretical uncertainty, such as the higher 
order coupled perturbation of vacuum
polarization and self-energy terms, quadruply excited cluster operators, etc.
However, in general, these all have much lower contributions to the properties 
and their cumulative theoretical uncertainty could be below 0.1\%.
The uncertainty in the wavelengths is estimated using the relative errors 
in the excitation energies of $^3P_1$ and $^1P_1$ states. 
The largest error 
is 2.4\% in the case of $^1P_1$. We choose this as an upper bound to the
uncertainty in wavelengths. Combining all sources of uncertainties, we 
get upper bound to the uncertainties in transition rates and HFS 
constants as 3\% and 1.6\%, respectively. The upper bound to the 
uncertainty in our computed $\alpha$ is about 3\% \cite{ravi-21b}.

\section{Conclusion} 

We have employed an all-particle FSRCC theory for two-valence atoms to 
investigate the ionization potential, excitation energies, transition rates
and HFS constants in superheavy nobelium. We combined these precision calculations
with available experimental data to extract the nuclear properties 
such as nuclear magnetic dipole and electric quadrupole moments. We also
employed a PRCC theory to compute the ground state electric dipole 
polarizability of No. To assess the accuracy of FSRCC and PRCC results, we
computed the ionization potential and dipole polarizability of 
lighter homolog Yb. In addition, to assess the nuclear deformation of
even-mass isotopes, we performed isotope shift calculations using MCDF
theory. To ensure the convergence of our FSRCC and PRCC results, we 
have employed large basis sets in the calculations. Moreover, to further 
improve the accuracy of our results, we incorporated the 
corrections from the Breit, QED and perturbative triples to
our calculations.

Our calculated IP is in good agreement with experimental 
data for both Yb and No, demonstrating the accuracy of FSRCC many-body 
wavefunction. The inclusion of high-energy two-valence configurations 
in the model space was observed to increase the accuracy of IP for both the 
systems due to accurate treatment of {\em valence-valence} electron 
correlation. Our computed transition rate for 
$^1S_{0}$ $\rightarrow$ $^1P_{1}$  transition is within 
the experimental error bar \cite{laatiaoui-16}. Our extracted values of
$\mu$ and $Q$ for No are in good agreement with CI+all-order 
calculation \cite{raeder-18}, however, with a small difference 
due to more accurate treatment of electron correlation effects in FSRCC theory.
Our extracted change in mean square charge radii of $^{252, 253, 255}$No 
isotopes is consistent with the previous 
theory calculations \cite{raeder-18, warbinek-24}. Our recommended value of ground 
state $\alpha$ for Yb is within the experimental error 
bar \cite{beloy-12, lei-15}. And for No, it is consistent with 
the previous CC calculations \cite{thierfelder-09}.

Perturbative triples are observed to contribute significantly to 
the properties. The largest contribution is found to be $\approx$ 10\%
in the case of transition rate for $^1S_{0}$ $\rightarrow$ $^3P_{1}$ 
transition. The largest combined Breit and QED contribution is observed 
to be $\approx$ 4\% in the case of HFS constant $A$ for $^1P_{1}$ 
state of No. The combined contribution to $\alpha$ from Breit+QED 
is observed to be 0.46\% and 0.26\% for Yb and No, respectively.

\begin{acknowledgments}
We would like to thank Suraj Pandey for useful discussions. One of the authors, 
BKM, acknowledges the funding support from SERB, DST (Grant No. CRG/2022/003845). 
Calculations are performed using the High Performance Computing cluster 
Tejas at the Indian Institute of Technology Delhi and PARAM Rudra, a national 
supercomputing facility at Inter-University Accelerator Centre (IUAC) New Delhi.
\end{acknowledgments}

\appendix

\section{Single-electron energies}

In the Table \ref{ene-no}, we provide the single-electron energies for Yb 
and No using GTOs and compared with the numerical data calculated from GRASP2K 
\cite{jonsson-13} and from the B-spline \cite{zatsarinny-16} basis. We have used
a $V_{(n-2)}$ potential to generate the GTO basis.

\begin{table*}[h]
        \caption{Orbital energies for core orbitals (in hartree) from GTO 
         is compared with the GRASP2K and B-spline energies for Yb and No. } 
        \label{ene-no}
        \begin{ruledtabular}
        \begin{tabular}{lcccccc}
         Orbitals  & \multicolumn{3}{c}{\rm Yb}  & \multicolumn{3}{c}{\rm No}  \\
  \cline{2-4} \cline{5-7} 
  & GTO   & B-spline  &  GRASP2K  & GTO & B-spline & GRASP2K \\
 \hline
    
          $1s_{1/2}$ & 2268.17773 & 2268.16522 & 2268.17808  & 5527.23105 &  5527.23232 & 5527.23029  \\
          $2s_{1/2}$ &  389.41778 &  389.41773 &  389.41819  & 1083.36599 &  1083.37585 & 1083.36533  \\
          $3s_{1/2}$ &   90.23285 &   90.23185 &   90.23327  &  285.94343 &   285.94493 &  285.94285  \\
          $4s_{1/2}$ &   19.19557 &   19.19543 &   19.19587  &   79.15814 &    79.15875 &   79.15768  \\
          $5s_{1/2}$ &    2.95992 &    2.95990 &    2.95996  &   19.34264 &    19.34281 &   19.34231  \\
          $6s_{1/2}$ &            &            &             &    3.33048 &     3.33054 &    3.33036  \\ 
                                  \hline 
          $2p_{1/2}$ &  370.58145 &  370.58423 &  370.58186  & 1047.90501 &  1047.90702 & 1047.90436  \\
          $3p_{1/2}$ &   81.94637 &   81.94639 &   81.94679  &  269.64229 &   269.64185 &  269.64166  \\
          $4p_{1/2}$ &   15.79963 &   15.79964 &   15.79991  &   71.50927 &    71.50933 &   71.50879  \\
          $5p_{1/2}$ &    1.94174 &    1.94174 &    1.94176  &   16.08047 &    16.08049 &   16.08014  \\
          $6p_{1/2}$ &            &            &             &    2.26509 &     2.26511 &    2.26518  \\
                                  \hline
          $2p_{3/2}$ &  332.01351 &  332.01641 &  332.01393  &  809.32579 &   809.32659 &  809.32517  \\
          $3p_{3/2}$ &   73.61816 &   73.61819 &   73.61859  &  212.72473 &   212.72399 &  212.72412  \\
          $4p_{3/2}$ &   13.89850 &   13.89852 &   13.89877  &   55.74889 &    55.74886 &   55.74843  \\
          $5p_{3/2}$ &    1.70394 &    1.70395 &    1.70397  &   11.97549 &    11.97553 &   11.97525  \\
          $6p_{3/2}$ &            &            &             &    1.58214 &     1.58215 &    1.58215  \\ 
                                  \hline 
          $3d_{3/2}$ &   59.71638 &   59.71639 &   59.71678  &  187.67181 &   187.67094 &  187.67107  \\
          $4d_{3/2}$ &    8.30308 &    8.30309 &    8.30329  &   43.77489 &    43.77471 &   43.77426  \\
          $5d_{3/2}$ &            &            &             &    7.14868 &     7.14868 &    7.14838  \\ 
                                  \hline 
          $3d_{5/2}$ &   57.91511 &   57.91512 &   57.91552  &  176.98193  &  176.98107 &  176.98119  \\
          $4d_{5/2}$ &    7.94745 &    7.94746 &    7.94765  &   40.99491  &   40.99474 &   40.99428  \\
          $5d_{5/2}$ &            &            &             &    6.52468  &    6.52469 &    6.52446  \\ 
                                  \hline 
          $4f_{5/2}$ &    1.06465 &    1.06458 &    1.06466  &   25.22607  &   25.22574 &   25.22532  \\
          $5f_{5/2}$ &            &            &             &    1.10555  &    1.10553 &    1.10539   \\ 
                                  \hline 
          $4f_{7/2}$ &    1.00614 &    1.00614 &    1.00596  &   24.45279  &   24.45246 &   24.45207  \\
          $5f_{7/2}$ &            &            &             &    1.00865  &    1.00864 &    1.00874  \\ 
                                \hline
          $E_{SCF}$  & 14067.06708  & 14067.01768 & 14067.06741 & 36740.15589 & 36740.28498 & 36740.16137  \\
        \end{tabular}  
        \end{ruledtabular}
\end{table*}                          


\bibliography{references}

\begin{thebibliography}{70}%
\makeatletter
\providecommand \@ifxundefined [1]{%
 \@ifx{#1\undefined}
}%
\providecommand \@ifnum [1]{%
 \ifnum #1\expandafter \@firstoftwo
 \else \expandafter \@secondoftwo
 \fi
}%
\providecommand \@ifx [1]{%
 \ifx #1\expandafter \@firstoftwo
 \else \expandafter \@secondoftwo
 \fi
}%
\providecommand \natexlab [1]{#1}%
\providecommand \enquote  [1]{``#1''}%
\providecommand \bibnamefont  [1]{#1}%
\providecommand \bibfnamefont [1]{#1}%
\providecommand \citenamefont [1]{#1}%
\providecommand \href@noop [0]{\@secondoftwo}%
\providecommand \href [0]{\begingroup \@sanitize@url \@href}%
\providecommand \@href[1]{\@@startlink{#1}\@@href}%
\providecommand \@@href[1]{\endgroup#1\@@endlink}%
\providecommand \@sanitize@url [0]{\catcode `\\12\catcode `\$12\catcode
  `\&12\catcode `\#12\catcode `\^12\catcode `\_12\catcode `\%12\relax}%
\providecommand \@@startlink[1]{}%
\providecommand \@@endlink[0]{}%
\providecommand \url  [0]{\begingroup\@sanitize@url \@url }%
\providecommand \@url [1]{\endgroup\@href {#1}{\urlprefix }}%
\providecommand \urlprefix  [0]{URL }%
\providecommand \Eprint [0]{\href }%
\providecommand \doibase [0]{https://doi.org/}%
\providecommand \selectlanguage [0]{\@gobble}%
\providecommand \bibinfo  [0]{\@secondoftwo}%
\providecommand \bibfield  [0]{\@secondoftwo}%
\providecommand \translation [1]{[#1]}%
\providecommand \BibitemOpen [0]{}%
\providecommand \bibitemStop [0]{}%
\providecommand \bibitemNoStop [0]{.\EOS\space}%
\providecommand \EOS [0]{\spacefactor3000\relax}%
\providecommand \BibitemShut  [1]{\csname bibitem#1\endcsname}%
\let\auto@bib@innerbib\@empty
\bibitem [{\citenamefont {Raeder~{\em et al.}}()}]{raeder-18}%
  \BibitemOpen
  \bibfield  {author} {\bibinfo {author} {\bibfnamefont {S.}~\bibnamefont
  {Raeder~{\em et al.}}},\ }\bibfield  {title} {\bibinfo {title} {Probing sizes
  and shapes of nobelium isotopes by laser spectroscopy},\ }\href@noop {} {\
  }\BibitemShut {NoStop}%
\bibitem [{\citenamefont {Türler}\ and\ \citenamefont
  {Pershina}(2013)}]{turler-13}%
  \BibitemOpen
  \bibfield  {author} {\bibinfo {author} {\bibfnamefont {A.}~\bibnamefont
  {Türler}}\ and\ \bibinfo {author} {\bibfnamefont {V.}~\bibnamefont
  {Pershina}},\ }\bibfield  {title} {\bibinfo {title} {Advances in the
  production and chemistry of the heaviest elements},\ }\href
  {https://doi.org/10.1021/cr3002438} {\bibfield  {journal} {\bibinfo
  {journal} {Chemical Reviews}\ }\textbf {\bibinfo {volume} {113}},\ \bibinfo
  {pages} {1237} (\bibinfo {year} {2013})},\ \bibinfo {note} {pMID: 23402305},\
  \Eprint {https://arxiv.org/abs/https://doi.org/10.1021/cr3002438}
  {https://doi.org/10.1021/cr3002438} \BibitemShut {NoStop}%
\bibitem [{\citenamefont {Shaughnessy}\ and\ \citenamefont
  {Schadel}(2014)}]{schadel-14}%
  \BibitemOpen
  \bibfield  {author} {\bibinfo {author} {\bibfnamefont {D.}~\bibnamefont
  {Shaughnessy}}\ and\ \bibinfo {author} {\bibfnamefont {M.}~\bibnamefont
  {Schadel}},\ }\href@noop {} {\emph {\bibinfo {title} {The Chemistry of the
  Superheavy Elements}}},\ \bibinfo {edition} {2nd}\ ed.\ (\bibinfo
  {publisher} {Springer},\ \bibinfo {address} {Heidelberg},\ \bibinfo {year}
  {2014})\BibitemShut {NoStop}%
\bibitem [{\citenamefont {Pershina}(2015)}]{pershina-15}%
  \BibitemOpen
  \bibfield  {author} {\bibinfo {author} {\bibfnamefont {V.}~\bibnamefont
  {Pershina}},\ }\bibfield  {title} {\bibinfo {title} {Electronic structure and
  properties of superheavy elements},\ }\href
  {https://doi.org/https://doi.org/10.1016/j.nuclphysa.2015.04.007} {\bibfield
  {journal} {\bibinfo  {journal} {Nuclear Physics A}\ }\textbf {\bibinfo
  {volume} {944}},\ \bibinfo {pages} {578 } (\bibinfo {year} {2015})},\
  \bibinfo {note} {special Issue on Superheavy Elements}\BibitemShut {NoStop}%
\bibitem [{\citenamefont {Schwerdtfeger}\ \emph {et~al.}(2015)\citenamefont
  {Schwerdtfeger}, \citenamefont {Pašteka}, \citenamefont {Punnett},\ and\
  \citenamefont {Bowman}}]{peter-15}%
  \BibitemOpen
  \bibfield  {author} {\bibinfo {author} {\bibfnamefont {P.}~\bibnamefont
  {Schwerdtfeger}}, \bibinfo {author} {\bibfnamefont {L.~F.}\ \bibnamefont
  {Pašteka}}, \bibinfo {author} {\bibfnamefont {A.}~\bibnamefont {Punnett}},\
  and\ \bibinfo {author} {\bibfnamefont {P.~O.}\ \bibnamefont {Bowman}},\
  }\bibfield  {title} {\bibinfo {title} {Relativistic and quantum
  electrodynamic effects in superheavy elements},\ }\href
  {https://doi.org/https://doi.org/10.1016/j.nuclphysa.2015.02.005} {\bibfield
  {journal} {\bibinfo  {journal} {Nuclear Physics A}\ }\textbf {\bibinfo
  {volume} {944}},\ \bibinfo {pages} {551 } (\bibinfo {year} {2015})},\
  \bibinfo {note} {special Issue on Superheavy Elements}\BibitemShut {NoStop}%
\bibitem [{\citenamefont {Eliav}\ \emph {et~al.}(2015)\citenamefont {Eliav},
  \citenamefont {Fritzsche},\ and\ \citenamefont {Kaldor}}]{eliav-15}%
  \BibitemOpen
  \bibfield  {author} {\bibinfo {author} {\bibfnamefont {E.}~\bibnamefont
  {Eliav}}, \bibinfo {author} {\bibfnamefont {S.}~\bibnamefont {Fritzsche}},\
  and\ \bibinfo {author} {\bibfnamefont {U.}~\bibnamefont {Kaldor}},\
  }\bibfield  {title} {\bibinfo {title} {Electronic structure theory of the
  superheavy elements},\ }\href
  {https://doi.org/https://doi.org/10.1016/j.nuclphysa.2015.06.017} {\bibfield
  {journal} {\bibinfo  {journal} {Nuclear Physics A}\ }\textbf {\bibinfo
  {volume} {944}},\ \bibinfo {pages} {518 } (\bibinfo {year} {2015})},\
  \bibinfo {note} {special Issue on Superheavy Elements}\BibitemShut {NoStop}%
\bibitem [{\citenamefont {Giuliani}\ \emph {et~al.}(2019)\citenamefont
  {Giuliani}, \citenamefont {Matheson}, \citenamefont {Nazarewicz},
  \citenamefont {Olsen}, \citenamefont {Reinhard}, \citenamefont {Sadhukhan},
  \citenamefont {Schuetrumpf}, \citenamefont {Schunck},\ and\ \citenamefont
  {Schwerdtfeger}}]{giuliani-19}%
  \BibitemOpen
  \bibfield  {author} {\bibinfo {author} {\bibfnamefont {S.~A.}\ \bibnamefont
  {Giuliani}}, \bibinfo {author} {\bibfnamefont {Z.}~\bibnamefont {Matheson}},
  \bibinfo {author} {\bibfnamefont {W.}~\bibnamefont {Nazarewicz}}, \bibinfo
  {author} {\bibfnamefont {E.}~\bibnamefont {Olsen}}, \bibinfo {author}
  {\bibfnamefont {P.-G.}\ \bibnamefont {Reinhard}}, \bibinfo {author}
  {\bibfnamefont {J.}~\bibnamefont {Sadhukhan}}, \bibinfo {author}
  {\bibfnamefont {B.}~\bibnamefont {Schuetrumpf}}, \bibinfo {author}
  {\bibfnamefont {N.}~\bibnamefont {Schunck}},\ and\ \bibinfo {author}
  {\bibfnamefont {P.}~\bibnamefont {Schwerdtfeger}},\ }\bibfield  {title}
  {\bibinfo {title} {Colloquium: Superheavy elements: Oganesson and beyond},\
  }\href {https://doi.org/10.1103/RevModPhys.91.011001} {\bibfield  {journal}
  {\bibinfo  {journal} {Rev. Mod. Phys.}\ }\textbf {\bibinfo {volume} {91}},\
  \bibinfo {pages} {011001} (\bibinfo {year} {2019})}\BibitemShut {NoStop}%
\bibitem [{\citenamefont {Pershina}\ and\ \citenamefont
  {Hoffman}(2008)}]{hoffman-08}%
  \BibitemOpen
  \bibfield  {author} {\bibinfo {author} {\bibfnamefont {V.}~\bibnamefont
  {Pershina}}\ and\ \bibinfo {author} {\bibfnamefont {D.}~\bibnamefont
  {Hoffman}},\ }\href@noop {} {\emph {\bibinfo {title} {Transactinide Elements
  and Future Elements}}}\ (\bibinfo  {publisher} {Springer, Dordrecht},\
  \bibinfo {year} {2008})\BibitemShut {NoStop}%
\bibitem [{\citenamefont {Laatiaoui~{\em et al.}}(2016)}]{laatiaoui-16}%
  \BibitemOpen
  \bibfield  {author} {\bibinfo {author} {\bibfnamefont {M.}~\bibnamefont
  {Laatiaoui~{\em et al.}}},\ }\bibfield  {title} {\bibinfo {title}
  {Atom-at-a-time laser resonance ionization spectroscopy of nobelium},\ }\href
  {https://doi.org/10.1038/nature19345} {\bibfield  {journal} {\bibinfo
  {journal} {Nature}\ }\textbf {\bibinfo {volume} {538}},\ \bibinfo {pages}
  {495} (\bibinfo {year} {2016})}\BibitemShut {NoStop}%
\bibitem [{\citenamefont {Chhetri~{\em et al.}}(2018)}]{chhetri-18}%
  \BibitemOpen
  \bibfield  {author} {\bibinfo {author} {\bibfnamefont {P.}~\bibnamefont
  {Chhetri~{\em et al.}}},\ }\bibfield  {title} {\bibinfo {title} {Precision
  measurement of the first ionization potential of nobelium},\ }\href
  {https://doi.org/10.1103/PhysRevLett.120.263003} {\bibfield  {journal}
  {\bibinfo  {journal} {Phys. Rev. Lett.}\ }\textbf {\bibinfo {volume} {120}},\
  \bibinfo {pages} {263003} (\bibinfo {year} {2018})}\BibitemShut {NoStop}%
\bibitem [{\citenamefont {Warbinek}\ \emph {et~al.}(2024)\citenamefont
  {Warbinek}, \citenamefont {Rickert}, \citenamefont {Raeder}, \citenamefont
  {Albrecht-Sch{\"o}nzart}, \citenamefont {Andelic}, \citenamefont {Auler},
  \citenamefont {Bally}, \citenamefont {Bender}, \citenamefont {Berndt},
  \citenamefont {Block} \emph {et~al.}}]{warbinek-24}%
  \BibitemOpen
  \bibfield  {author} {\bibinfo {author} {\bibfnamefont {J.}~\bibnamefont
  {Warbinek}}, \bibinfo {author} {\bibfnamefont {E.}~\bibnamefont {Rickert}},
  \bibinfo {author} {\bibfnamefont {S.}~\bibnamefont {Raeder}}, \bibinfo
  {author} {\bibfnamefont {T.}~\bibnamefont {Albrecht-Sch{\"o}nzart}}, \bibinfo
  {author} {\bibfnamefont {B.}~\bibnamefont {Andelic}}, \bibinfo {author}
  {\bibfnamefont {J.}~\bibnamefont {Auler}}, \bibinfo {author} {\bibfnamefont
  {B.}~\bibnamefont {Bally}}, \bibinfo {author} {\bibfnamefont
  {M.}~\bibnamefont {Bender}}, \bibinfo {author} {\bibfnamefont
  {S.}~\bibnamefont {Berndt}}, \bibinfo {author} {\bibfnamefont
  {M.}~\bibnamefont {Block}}, \emph {et~al.},\ }\bibfield  {title} {\bibinfo
  {title} {Smooth trends in fermium charge radii and the impact of shell
  effects},\ }\href {https://doi.org/10.1038/s41586-024-08062-z} {\bibfield
  {journal} {\bibinfo  {journal} {Nature}\ }\textbf {\bibinfo {volume} {634}},\
  \bibinfo {pages} {1075} (\bibinfo {year} {2024})}\BibitemShut {NoStop}%
\bibitem [{\citenamefont {Liu}\ \emph {et~al.}(2007)\citenamefont {Liu},
  \citenamefont {Hutton},\ and\ \citenamefont {Zou}}]{liu-07}%
  \BibitemOpen
  \bibfield  {author} {\bibinfo {author} {\bibfnamefont {Y.}~\bibnamefont
  {Liu}}, \bibinfo {author} {\bibfnamefont {R.}~\bibnamefont {Hutton}},\ and\
  \bibinfo {author} {\bibfnamefont {Y.}~\bibnamefont {Zou}},\ }\bibfield
  {title} {\bibinfo {title} {Atomic structure of the super-heavy element
  $\mathrm{No}$ $\mathrm{I}$ $(\mathrm{Z}=102)$},\ }\href
  {https://doi.org/10.1103/PhysRevA.76.062503} {\bibfield  {journal} {\bibinfo
  {journal} {Phys. Rev. A}\ }\textbf {\bibinfo {volume} {76}},\ \bibinfo
  {pages} {062503} (\bibinfo {year} {2007})}\BibitemShut {NoStop}%
\bibitem [{\citenamefont {Borschevsky}\ \emph {et~al.}(2007)\citenamefont
  {Borschevsky}, \citenamefont {Eliav}, \citenamefont {Vilkas}, \citenamefont
  {Ishikawa},\ and\ \citenamefont {Kaldor}}]{borschevsky-07}%
  \BibitemOpen
  \bibfield  {author} {\bibinfo {author} {\bibfnamefont {A.}~\bibnamefont
  {Borschevsky}}, \bibinfo {author} {\bibfnamefont {E.}~\bibnamefont {Eliav}},
  \bibinfo {author} {\bibfnamefont {M.~J.}\ \bibnamefont {Vilkas}}, \bibinfo
  {author} {\bibfnamefont {Y.}~\bibnamefont {Ishikawa}},\ and\ \bibinfo
  {author} {\bibfnamefont {U.}~\bibnamefont {Kaldor}},\ }\bibfield  {title}
  {\bibinfo {title} {Predicted spectrum of atomic nobelium},\ }\href
  {https://doi.org/10.1103/PhysRevA.75.042514} {\bibfield  {journal} {\bibinfo
  {journal} {Phys. Rev. A}\ }\textbf {\bibinfo {volume} {75}},\ \bibinfo
  {pages} {042514} (\bibinfo {year} {2007})}\BibitemShut {NoStop}%
\bibitem [{\citenamefont {Indelicato}\ \emph {et~al.}(2007)\citenamefont
  {Indelicato}, \citenamefont {Santos}, \citenamefont {Boucard},\ and\
  \citenamefont {Desclaux}}]{indelicato-07}%
  \BibitemOpen
  \bibfield  {author} {\bibinfo {author} {\bibfnamefont {P.}~\bibnamefont
  {Indelicato}}, \bibinfo {author} {\bibfnamefont {J.}~\bibnamefont {Santos}},
  \bibinfo {author} {\bibfnamefont {S.}~\bibnamefont {Boucard}},\ and\ \bibinfo
  {author} {\bibfnamefont {J.-P.}\ \bibnamefont {Desclaux}},\ }\bibfield
  {title} {\bibinfo {title} {Qed and relativistic corrections in superheavy
  elements},\ }\href {https://doi.org/10.1140/epjd/e2007-00229-y} {\bibfield
  {journal} {\bibinfo  {journal} {The European Physical Journal D}\ }\textbf
  {\bibinfo {volume} {45}},\ \bibinfo {pages} {155} (\bibinfo {year}
  {2007})}\BibitemShut {NoStop}%
\bibitem [{\citenamefont {Thierfelder}\ and\ \citenamefont
  {Schwerdtfeger}(2009{\natexlab{a}})}]{thierfelder-09}%
  \BibitemOpen
  \bibfield  {author} {\bibinfo {author} {\bibfnamefont {C.}~\bibnamefont
  {Thierfelder}}\ and\ \bibinfo {author} {\bibfnamefont {P.}~\bibnamefont
  {Schwerdtfeger}},\ }\bibfield  {title} {\bibinfo {title} {Effect of
  relativity and electron correlation in static dipole polarizabilities of
  ytterbium and nobelium},\ }\href {https://doi.org/10.1103/PhysRevA.79.032512}
  {\bibfield  {journal} {\bibinfo  {journal} {Phys. Rev. A}\ }\textbf {\bibinfo
  {volume} {79}},\ \bibinfo {pages} {032512} (\bibinfo {year}
  {2009}{\natexlab{a}})}\BibitemShut {NoStop}%
\bibitem [{\citenamefont {Safronova}\ \emph {et~al.}(1999)\citenamefont
  {Safronova}, \citenamefont {Johnson},\ and\ \citenamefont
  {Derevianko}}]{safronova-99}%
  \BibitemOpen
  \bibfield  {author} {\bibinfo {author} {\bibfnamefont {M.~S.}\ \bibnamefont
  {Safronova}}, \bibinfo {author} {\bibfnamefont {W.~R.}\ \bibnamefont
  {Johnson}},\ and\ \bibinfo {author} {\bibfnamefont {A.}~\bibnamefont
  {Derevianko}},\ }\bibfield  {title} {\bibinfo {title} {Relativistic many-body
  calculations of energy levels, hyperfine constants, electric-dipole matrix
  elements, and static polarizabilities for alkali-metal atoms},\ }\href
  {https://doi.org/10.1103/PhysRevA.60.4476} {\bibfield  {journal} {\bibinfo
  {journal} {Phys. Rev. A}\ }\textbf {\bibinfo {volume} {60}},\ \bibinfo
  {pages} {4476} (\bibinfo {year} {1999})}\BibitemShut {NoStop}%
\bibitem [{\citenamefont {Derevianko}\ \emph {et~al.}(1999)\citenamefont
  {Derevianko}, \citenamefont {Johnson}, \citenamefont {Safronova},\ and\
  \citenamefont {Babb}}]{derevianko-99}%
  \BibitemOpen
  \bibfield  {author} {\bibinfo {author} {\bibfnamefont {A.}~\bibnamefont
  {Derevianko}}, \bibinfo {author} {\bibfnamefont {W.~R.}\ \bibnamefont
  {Johnson}}, \bibinfo {author} {\bibfnamefont {M.~S.}\ \bibnamefont
  {Safronova}},\ and\ \bibinfo {author} {\bibfnamefont {J.~F.}\ \bibnamefont
  {Babb}},\ }\bibfield  {title} {\bibinfo {title} {High-precision calculations
  of dispersion coefficients, static dipole polarizabilities, and atom-wall
  interaction constants for alkali-metal atoms},\ }\href
  {https://doi.org/10.1103/PhysRevLett.82.3589} {\bibfield  {journal} {\bibinfo
   {journal} {Phys. Rev. Lett.}\ }\textbf {\bibinfo {volume} {82}},\ \bibinfo
  {pages} {3589} (\bibinfo {year} {1999})}\BibitemShut {NoStop}%
\bibitem [{\citenamefont {Chattopadhyay}\ \emph
  {et~al.}(2012{\natexlab{a}})\citenamefont {Chattopadhyay}, \citenamefont
  {Mani},\ and\ \citenamefont {Angom}}]{chattopadhyay-12b}%
  \BibitemOpen
  \bibfield  {author} {\bibinfo {author} {\bibfnamefont {S.}~\bibnamefont
  {Chattopadhyay}}, \bibinfo {author} {\bibfnamefont {B.~K.}\ \bibnamefont
  {Mani}},\ and\ \bibinfo {author} {\bibfnamefont {D.}~\bibnamefont {Angom}},\
  }\bibfield  {title} {\bibinfo {title} {Perturbed coupled-cluster theory to
  calculate dipole polarizabilities of closed-shell systems: Application to
  {Ar, Kr, Xe, and Rn}},\ }\href {https://doi.org/10.1103/PhysRevA.86.062508}
  {\bibfield  {journal} {\bibinfo  {journal} {Phys. Rev. A}\ }\textbf {\bibinfo
  {volume} {86}},\ \bibinfo {pages} {062508} (\bibinfo {year}
  {2012}{\natexlab{a}})}\BibitemShut {NoStop}%
\bibitem [{\citenamefont {Chattopadhyay}\ \emph {et~al.}(2013)\citenamefont
  {Chattopadhyay}, \citenamefont {Mani},\ and\ \citenamefont
  {Angom}}]{chattopadhyay-13b}%
  \BibitemOpen
  \bibfield  {author} {\bibinfo {author} {\bibfnamefont {S.}~\bibnamefont
  {Chattopadhyay}}, \bibinfo {author} {\bibfnamefont {B.~K.}\ \bibnamefont
  {Mani}},\ and\ \bibinfo {author} {\bibfnamefont {D.}~\bibnamefont {Angom}},\
  }\bibfield  {title} {\bibinfo {title} {Electric dipole polarizabilities of
  doubly ionized alkaline-earth-metal ions from perturbed relativistic
  coupled-cluster theory},\ }\href {https://doi.org/10.1103/PhysRevA.87.062504}
  {\bibfield  {journal} {\bibinfo  {journal} {Phys. Rev. A}\ }\textbf {\bibinfo
  {volume} {87}},\ \bibinfo {pages} {062504} (\bibinfo {year}
  {2013})}\BibitemShut {NoStop}%
\bibitem [{\citenamefont {Chattopadhyay}\ \emph {et~al.}(2014)\citenamefont
  {Chattopadhyay}, \citenamefont {Mani},\ and\ \citenamefont
  {Angom}}]{chattopadhyay-14}%
  \BibitemOpen
  \bibfield  {author} {\bibinfo {author} {\bibfnamefont {S.}~\bibnamefont
  {Chattopadhyay}}, \bibinfo {author} {\bibfnamefont {B.~K.}\ \bibnamefont
  {Mani}},\ and\ \bibinfo {author} {\bibfnamefont {D.}~\bibnamefont {Angom}},\
  }\bibfield  {title} {\bibinfo {title} {Electric dipole polarizability of
  alkaline-earth-metal atoms from perturbed relativistic coupled-cluster theory
  with triples},\ }\href {https://doi.org/10.1103/PhysRevA.89.022506}
  {\bibfield  {journal} {\bibinfo  {journal} {Phys. Rev. A}\ }\textbf {\bibinfo
  {volume} {89}},\ \bibinfo {pages} {022506} (\bibinfo {year}
  {2014})}\BibitemShut {NoStop}%
\bibitem [{\citenamefont {Chattopadhyay}\ \emph {et~al.}(2015)\citenamefont
  {Chattopadhyay}, \citenamefont {Mani},\ and\ \citenamefont
  {Angom}}]{chattopadhyay-15}%
  \BibitemOpen
  \bibfield  {author} {\bibinfo {author} {\bibfnamefont {S.}~\bibnamefont
  {Chattopadhyay}}, \bibinfo {author} {\bibfnamefont {B.~K.}\ \bibnamefont
  {Mani}},\ and\ \bibinfo {author} {\bibfnamefont {D.}~\bibnamefont {Angom}},\
  }\bibfield  {title} {\bibinfo {title} {Triple excitations in perturbed
  relativistic coupled-cluster theory and electric dipole polarizability of
  group{IIB} elements},\ }\href
  {https://link.aps.org/doi/10.1103/PhysRevA.91.052504} {\bibfield  {journal}
  {\bibinfo  {journal} {Phys. Rev. A}\ }\textbf {\bibinfo {volume} {91}},\
  \bibinfo {pages} {052504} (\bibinfo {year} {2015})}\BibitemShut {NoStop}%
\bibitem [{\citenamefont {Kumar}\ \emph {et~al.}(2020)\citenamefont {Kumar},
  \citenamefont {Chattopadhyay}, \citenamefont {Mani},\ and\ \citenamefont
  {Angom}}]{ravi-20}%
  \BibitemOpen
  \bibfield  {author} {\bibinfo {author} {\bibfnamefont {R.}~\bibnamefont
  {Kumar}}, \bibinfo {author} {\bibfnamefont {S.}~\bibnamefont
  {Chattopadhyay}}, \bibinfo {author} {\bibfnamefont {B.~K.}\ \bibnamefont
  {Mani}},\ and\ \bibinfo {author} {\bibfnamefont {D.}~\bibnamefont {Angom}},\
  }\bibfield  {title} {\bibinfo {title} {Electric dipole polarizability of
  group-13 ions using perturbed relativistic coupled-cluster theory: Importance
  of nonlinear terms},\ }\href {https://doi.org/10.1103/PhysRevA.101.012503}
  {\bibfield  {journal} {\bibinfo  {journal} {Phys. Rev. A}\ }\textbf {\bibinfo
  {volume} {101}},\ \bibinfo {pages} {012503} (\bibinfo {year}
  {2020})}\BibitemShut {NoStop}%
\bibitem [{\citenamefont {Mani}\ and\ \citenamefont {Angom}(2011)}]{mani-11}%
  \BibitemOpen
  \bibfield  {author} {\bibinfo {author} {\bibfnamefont {B.~K.}\ \bibnamefont
  {Mani}}\ and\ \bibinfo {author} {\bibfnamefont {D.}~\bibnamefont {Angom}},\
  }\bibfield  {title} {\bibinfo {title} {Fock-space relativistic
  coupled-cluster calculations of two-valence atoms},\ }\href
  {https://doi.org/10.1103/PhysRevA.83.012501} {\bibfield  {journal} {\bibinfo
  {journal} {Phys. Rev. A}\ }\textbf {\bibinfo {volume} {83}},\ \bibinfo
  {pages} {012501} (\bibinfo {year} {2011})}\BibitemShut {NoStop}%
\bibitem [{\citenamefont {Kumar}\ \emph
  {et~al.}(2021{\natexlab{a}})\citenamefont {Kumar}, \citenamefont
  {Chattopadhyay}, \citenamefont {Angom},\ and\ \citenamefont
  {Mani}}]{ravi-21a}%
  \BibitemOpen
  \bibfield  {author} {\bibinfo {author} {\bibfnamefont {R.}~\bibnamefont
  {Kumar}}, \bibinfo {author} {\bibfnamefont {S.}~\bibnamefont
  {Chattopadhyay}}, \bibinfo {author} {\bibfnamefont {D.}~\bibnamefont
  {Angom}},\ and\ \bibinfo {author} {\bibfnamefont {B.~K.}\ \bibnamefont
  {Mani}},\ }\bibfield  {title} {\bibinfo {title} {Fock-space relativistic
  coupled-cluster calculation of a hyperfine-induced
  ${}^{1}{S}_{0}\ensuremath{\rightarrow}{}^{3}{P}_{0}^{o}$ clock transition in
  {Al}$^+$},\ }\href {https://doi.org/10.1103/PhysRevA.103.022801} {\bibfield
  {journal} {\bibinfo  {journal} {Phys. Rev. A}\ }\textbf {\bibinfo {volume}
  {103}},\ \bibinfo {pages} {022801} (\bibinfo {year}
  {2021}{\natexlab{a}})}\BibitemShut {NoStop}%
\bibitem [{\citenamefont {Mani}\ \emph {et~al.}(2009)\citenamefont {Mani},
  \citenamefont {Latha},\ and\ \citenamefont {Angom}}]{mani-09}%
  \BibitemOpen
  \bibfield  {author} {\bibinfo {author} {\bibfnamefont {B.~K.}\ \bibnamefont
  {Mani}}, \bibinfo {author} {\bibfnamefont {K.~V.~P.}\ \bibnamefont {Latha}},\
  and\ \bibinfo {author} {\bibfnamefont {D.}~\bibnamefont {Angom}},\ }\bibfield
   {title} {\bibinfo {title} {Relativistic coupled-cluster calculations of
  $^{20}\text{N}\text{e}$, $^{40}\text{A}\text{r}$, $^{84}\text{K}\text{r}$,
  and $^{129}\text{X}\text{e}$: Correlation energies and dipole
  polarizabilities},\ }\href {https://doi.org/10.1103/PhysRevA.80.062505}
  {\bibfield  {journal} {\bibinfo  {journal} {Phys. Rev. A}\ }\textbf {\bibinfo
  {volume} {80}},\ \bibinfo {pages} {062505} (\bibinfo {year}
  {2009})}\BibitemShut {NoStop}%
\bibitem [{\citenamefont {Chattopadhyay}\ \emph
  {et~al.}(2012{\natexlab{b}})\citenamefont {Chattopadhyay}, \citenamefont
  {Mani},\ and\ \citenamefont {Angom}}]{chattopadhyay-12a}%
  \BibitemOpen
  \bibfield  {author} {\bibinfo {author} {\bibfnamefont {S.}~\bibnamefont
  {Chattopadhyay}}, \bibinfo {author} {\bibfnamefont {B.~K.}\ \bibnamefont
  {Mani}},\ and\ \bibinfo {author} {\bibfnamefont {D.}~\bibnamefont {Angom}},\
  }\bibfield  {title} {\bibinfo {title} {Electric dipole polarizability from
  perturbed relativistic coupled-cluster theory: Application to neon},\ }\href
  {https://doi.org/10.1103/PhysRevA.86.022522} {\bibfield  {journal} {\bibinfo
  {journal} {Phys. Rev. A}\ }\textbf {\bibinfo {volume} {86}},\ \bibinfo
  {pages} {022522} (\bibinfo {year} {2012}{\natexlab{b}})}\BibitemShut
  {NoStop}%
\bibitem [{\citenamefont {Kumar}\ \emph
  {et~al.}(2021{\natexlab{b}})\citenamefont {Kumar}, \citenamefont
  {Chattopadhyay}, \citenamefont {Angom},\ and\ \citenamefont
  {Mani}}]{ravi-21b}%
  \BibitemOpen
  \bibfield  {author} {\bibinfo {author} {\bibfnamefont {R.}~\bibnamefont
  {Kumar}}, \bibinfo {author} {\bibfnamefont {S.}~\bibnamefont
  {Chattopadhyay}}, \bibinfo {author} {\bibfnamefont {D.}~\bibnamefont
  {Angom}},\ and\ \bibinfo {author} {\bibfnamefont {B.~K.}\ \bibnamefont
  {Mani}},\ }\bibfield  {title} {\bibinfo {title} {Relativistic coupled-cluster
  calculation of the electric dipole polarizability and correlation energy of
  {Cn, ${\mathrm{Nh}}^{+}$, and Og}: Correlation effects from lighter to
  superheavy elements},\ }\href {https://doi.org/10.1103/PhysRevA.103.062803}
  {\bibfield  {journal} {\bibinfo  {journal} {Phys. Rev. A}\ }\textbf {\bibinfo
  {volume} {103}},\ \bibinfo {pages} {062803} (\bibinfo {year}
  {2021}{\natexlab{b}})}\BibitemShut {NoStop}%
\bibitem [{\citenamefont {Mani}\ and\ \citenamefont {Angom}(2010)}]{mani-10}%
  \BibitemOpen
  \bibfield  {author} {\bibinfo {author} {\bibfnamefont {B.~K.}\ \bibnamefont
  {Mani}}\ and\ \bibinfo {author} {\bibfnamefont {D.}~\bibnamefont {Angom}},\
  }\bibfield  {title} {\bibinfo {title} {Atomic properties calculated by
  relativistic coupled-cluster theory without truncation: Hyperfine constants
  of {${\mathrm{Mg}}^{+}$, ${\mathrm{Ca}}^{+}$, ${\mathrm{Sr}}^{+}$, and
  ${\mathrm{Ba}}^{+}$}},\ }\href {https://doi.org/10.1103/PhysRevA.81.042514}
  {\bibfield  {journal} {\bibinfo  {journal} {Phys. Rev. A}\ }\textbf {\bibinfo
  {volume} {81}},\ \bibinfo {pages} {042514} (\bibinfo {year}
  {2010})}\BibitemShut {NoStop}%
\bibitem [{\citenamefont {Mani}\ \emph {et~al.}(2017)\citenamefont {Mani},
  \citenamefont {Chattopadhyay},\ and\ \citenamefont {Angom}}]{mani-17}%
  \BibitemOpen
  \bibfield  {author} {\bibinfo {author} {\bibfnamefont {B.}~\bibnamefont
  {Mani}}, \bibinfo {author} {\bibfnamefont {S.}~\bibnamefont
  {Chattopadhyay}},\ and\ \bibinfo {author} {\bibfnamefont {D.}~\bibnamefont
  {Angom}},\ }\bibfield  {title} {\bibinfo {title} {Rccpac: A parallel
  relativistic coupled-cluster program for closed-shell and one-valence atoms
  and ions in fortran},\ }\href
  {https://doi.org/https://doi.org/10.1016/j.cpc.2016.11.008} {\bibfield
  {journal} {\bibinfo  {journal} {Computer Physics Communications}\ }\textbf
  {\bibinfo {volume} {213}},\ \bibinfo {pages} {136 } (\bibinfo {year}
  {2017})}\BibitemShut {NoStop}%
\bibitem [{\citenamefont {Gakkhar}\ \emph {et~al.}(2024)\citenamefont
  {Gakkhar}, \citenamefont {Kumar}, \citenamefont {Angom},\ and\ \citenamefont
  {Mani}}]{palki-24}%
  \BibitemOpen
  \bibfield  {author} {\bibinfo {author} {\bibfnamefont {P.}~\bibnamefont
  {Gakkhar}}, \bibinfo {author} {\bibfnamefont {R.}~\bibnamefont {Kumar}},
  \bibinfo {author} {\bibfnamefont {D.}~\bibnamefont {Angom}},\ and\ \bibinfo
  {author} {\bibfnamefont {B.~K.}\ \bibnamefont {Mani}},\ }\bibfield  {title}
  {\bibinfo {title} {Fock-space relativistic coupled-cluster calculations of
  clock-transition properties in {Pb}$^{2+}$},\ }\href
  {https://doi.org/10.1103/PhysRevA.110.013119} {\bibfield  {journal} {\bibinfo
   {journal} {Phys. Rev. A}\ }\textbf {\bibinfo {volume} {110}},\ \bibinfo
  {pages} {013119} (\bibinfo {year} {2024})}\BibitemShut {NoStop}%
\bibitem [{\citenamefont {Pulay}(1980)}]{pulay-80}%
  \BibitemOpen
  \bibfield  {author} {\bibinfo {author} {\bibfnamefont {P.}~\bibnamefont
  {Pulay}},\ }\bibfield  {title} {\bibinfo {title} {Convergence acceleration of
  iterative sequences. the case of scf iteration},\ }\href
  {https://doi.org/10.1016/0009-2614(80)80396-4} {\bibfield  {journal}
  {\bibinfo  {journal} {Chem. Phys. Lett.}\ }\textbf {\bibinfo {volume} {73}},\
  \bibinfo {pages} {393 } (\bibinfo {year} {1980})}\BibitemShut {NoStop}%
\bibitem [{\citenamefont {Mohanty}\ \emph {et~al.}(1991)\citenamefont
  {Mohanty}, \citenamefont {Parpia},\ and\ \citenamefont
  {Clementi}}]{mohanty-91}%
  \BibitemOpen
  \bibfield  {author} {\bibinfo {author} {\bibfnamefont {A.~K.}\ \bibnamefont
  {Mohanty}}, \bibinfo {author} {\bibfnamefont {F.~A.}\ \bibnamefont
  {Parpia}},\ and\ \bibinfo {author} {\bibfnamefont {E.}~\bibnamefont
  {Clementi}},\ }\bibfield  {title} {\bibinfo {title} {Kinetically balanced
  geometric gaussian basis set calculations for relativistic many-electron
  atoms},\ }in\ \href@noop {} {\emph {\bibinfo {booktitle} {Modern Techniques
  in Computational Chemistry: MOTECC-91}}},\ \bibinfo {editor} {edited by\
  \bibinfo {editor} {\bibfnamefont {E.}~\bibnamefont {Clementi}}}\ (\bibinfo
  {publisher} {ESCOM},\ \bibinfo {year} {1991})\BibitemShut {NoStop}%
\bibitem [{\citenamefont {J{\"{o}}nsson}\ \emph {et~al.}(2013)\citenamefont
  {J{\"{o}}nsson}, \citenamefont {Gaigalas}, \citenamefont {Biero{\'{n}}},
  \citenamefont {Froese~Fischer},\ and\ \citenamefont {Grant}}]{jonsson-13}%
  \BibitemOpen
  \bibfield  {author} {\bibinfo {author} {\bibfnamefont {P.}~\bibnamefont
  {J{\"{o}}nsson}}, \bibinfo {author} {\bibfnamefont {G.}~\bibnamefont
  {Gaigalas}}, \bibinfo {author} {\bibfnamefont {J.}~\bibnamefont
  {Biero{\'{n}}}}, \bibinfo {author} {\bibfnamefont {C.}~\bibnamefont
  {Froese~Fischer}},\ and\ \bibinfo {author} {\bibfnamefont {I.~P.}\
  \bibnamefont {Grant}},\ }\bibfield  {title} {\bibinfo {title} {New version:
  Grasp2k relativistic atomic structure package},\ }\href
  {https://doi.org/http://dx.doi.org/10.1016/j.cpc.2013.02.016} {\bibfield
  {journal} {\bibinfo  {journal} {Comp. Phys. Comm.}\ }\textbf {\bibinfo
  {volume} {184}},\ \bibinfo {pages} {2197 } (\bibinfo {year}
  {2013})}\BibitemShut {NoStop}%
\bibitem [{\citenamefont {Zatsarinny}\ and\ \citenamefont
  {Fischer}(2016)}]{oleg-16}%
  \BibitemOpen
  \bibfield  {author} {\bibinfo {author} {\bibfnamefont {O.}~\bibnamefont
  {Zatsarinny}}\ and\ \bibinfo {author} {\bibfnamefont {C.~F.}\ \bibnamefont
  {Fischer}},\ }\bibfield  {title} {\bibinfo {title} {{DBSR$_{\rm HF}$: A
  B-spline Dirac-Hartree-Fock program}},\ }\href
  {https://doi.org/https://doi.org/10.1016/j.cpc.2015.12.023} {\bibfield
  {journal} {\bibinfo  {journal} {Computer Physics Communications}\ }\textbf
  {\bibinfo {volume} {202}},\ \bibinfo {pages} {287 } (\bibinfo {year}
  {2016})}\BibitemShut {NoStop}%
\bibitem [{\citenamefont {Shabaev}\ \emph {et~al.}(2015)\citenamefont
  {Shabaev}, \citenamefont {Tupitsyn},\ and\ \citenamefont
  {Yerokhin}}]{shabaev-15}%
  \BibitemOpen
  \bibfield  {author} {\bibinfo {author} {\bibfnamefont {V.}~\bibnamefont
  {Shabaev}}, \bibinfo {author} {\bibfnamefont {I.}~\bibnamefont {Tupitsyn}},\
  and\ \bibinfo {author} {\bibfnamefont {V.}~\bibnamefont {Yerokhin}},\
  }\bibfield  {title} {\bibinfo {title} {Qedmod: Fortran program for
  calculating the model lamb-shift operator},\ }\href
  {https://doi.org/https://doi.org/10.1016/j.cpc.2014.12.002} {\bibfield
  {journal} {\bibinfo  {journal} {Computer Physics Communications}\ }\textbf
  {\bibinfo {volume} {189}},\ \bibinfo {pages} {175} (\bibinfo {year}
  {2015})}\BibitemShut {NoStop}%
\bibitem [{\citenamefont {Uehling}(1935)}]{uehling-35}%
  \BibitemOpen
  \bibfield  {author} {\bibinfo {author} {\bibfnamefont {E.~A.}\ \bibnamefont
  {Uehling}},\ }\bibfield  {title} {\bibinfo {title} {Polarization effects in
  the positron theory},\ }\href {https://doi.org/10.1103/PhysRev.48.55}
  {\bibfield  {journal} {\bibinfo  {journal} {Phys. Rev.}\ }\textbf {\bibinfo
  {volume} {48}},\ \bibinfo {pages} {55} (\bibinfo {year} {1935})}\BibitemShut
  {NoStop}%
\bibitem [{\citenamefont {Grant}(2006)}]{grant-06}%
  \BibitemOpen
  \bibfield  {author} {\bibinfo {author} {\bibfnamefont {I.}~\bibnamefont
  {Grant}},\ }\bibfield  {title} {\bibinfo {title} {Relativistic atomic
  structure},\ }in\ \href {https://doi.org/10.1007/978-0-387-26308-3_22} {\emph
  {\bibinfo {booktitle} {Springer Handbook of Atomic, Molecular, and Optical
  Physics}}},\ \bibinfo {editor} {edited by\ \bibinfo {editor} {\bibfnamefont
  {G.}~\bibnamefont {Drake}}}\ (\bibinfo  {publisher} {Springer},\ \bibinfo
  {address} {New York},\ \bibinfo {year} {2006})\ pp.\ \bibinfo {pages}
  {325--357}\BibitemShut {NoStop}%
\bibitem [{\citenamefont {Eliav}\ \emph {et~al.}(1995)\citenamefont {Eliav},
  \citenamefont {Kaldor},\ and\ \citenamefont {Ishikawa}}]{eliav-95}%
  \BibitemOpen
  \bibfield  {author} {\bibinfo {author} {\bibfnamefont {E.}~\bibnamefont
  {Eliav}}, \bibinfo {author} {\bibfnamefont {U.}~\bibnamefont {Kaldor}},\ and\
  \bibinfo {author} {\bibfnamefont {Y.}~\bibnamefont {Ishikawa}},\ }\bibfield
  {title} {\bibinfo {title} {Transition energies of ytterbium, lutetium, and
  lawrencium by the relativistic coupled-cluster method},\ }\href
  {https://doi.org/10.1103/PhysRevA.52.291} {\bibfield  {journal} {\bibinfo
  {journal} {Physical Review A}\ }\textbf {\bibinfo {volume} {52}},\ \bibinfo
  {pages} {291} (\bibinfo {year} {1995})}\BibitemShut {NoStop}%
\bibitem [{\citenamefont {Cao}\ and\ \citenamefont {Dolg}(2001)}]{cao-01}%
  \BibitemOpen
  \bibfield  {author} {\bibinfo {author} {\bibfnamefont {X.}~\bibnamefont
  {Cao}}\ and\ \bibinfo {author} {\bibfnamefont {M.}~\bibnamefont {Dolg}},\
  }\bibfield  {title} {\bibinfo {title} {Valence basis sets for relativistic
  energy-consistent small-core lanthanide pseudopotentials},\ }\href
  {https://doi.org/10.1063/1.1406535} {\bibfield  {journal} {\bibinfo
  {journal} {The Journal of Chemical Physics}\ }\textbf {\bibinfo {volume}
  {115}},\ \bibinfo {pages} {7348} (\bibinfo {year} {2001})}\BibitemShut
  {NoStop}%
\bibitem [{\citenamefont {{Nayak, Malaya K.}}\ and\ \citenamefont {{Chaudhuri,
  Rajat K.}}(2006)}]{nayak-06}%
  \BibitemOpen
  \bibfield  {author} {\bibinfo {author} {\bibnamefont {{Nayak, Malaya K.}}}\
  and\ \bibinfo {author} {\bibnamefont {{Chaudhuri, Rajat K.}}},\ }\bibfield
  {title} {\bibinfo {title} {Relativistic coupled cluster method - excitation
  and ionization energies of sr and yb atom},\ }\href
  {https://doi.org/10.1140/epjd/e2005-00279-1} {\bibfield  {journal} {\bibinfo
  {journal} {Eur. Phys. J. D}\ }\textbf {\bibinfo {volume} {37}},\ \bibinfo
  {pages} {171} (\bibinfo {year} {2006})}\BibitemShut {NoStop}%
\bibitem [{\citenamefont {Cao}\ \emph {et~al.}(2003)\citenamefont {Cao},
  \citenamefont {Dolg},\ and\ \citenamefont {Stoll}}]{cao-03}%
  \BibitemOpen
  \bibfield  {author} {\bibinfo {author} {\bibfnamefont {X.}~\bibnamefont
  {Cao}}, \bibinfo {author} {\bibfnamefont {M.}~\bibnamefont {Dolg}},\ and\
  \bibinfo {author} {\bibfnamefont {H.}~\bibnamefont {Stoll}},\ }\bibfield
  {title} {\bibinfo {title} {Valence basis sets for relativistic
  energy-consistent small-core actinide pseudopotentials},\ }\href
  {https://doi.org/10.1063/1.1521431} {\bibfield  {journal} {\bibinfo
  {journal} {The journal of chemical physics}\ }\textbf {\bibinfo {volume}
  {118}},\ \bibinfo {pages} {487} (\bibinfo {year} {2003})}\BibitemShut
  {NoStop}%
\bibitem [{\citenamefont {Dzuba}\ \emph
  {et~al.}(2014{\natexlab{a}})\citenamefont {Dzuba}, \citenamefont
  {Safronova},\ and\ \citenamefont {Safronova}}]{dzuba-14}%
  \BibitemOpen
  \bibfield  {author} {\bibinfo {author} {\bibfnamefont {V.}~\bibnamefont
  {Dzuba}}, \bibinfo {author} {\bibfnamefont {M.}~\bibnamefont {Safronova}},\
  and\ \bibinfo {author} {\bibfnamefont {U.}~\bibnamefont {Safronova}},\
  }\bibfield  {title} {\bibinfo {title} {Atomic properties of superheavy
  elements $\mathrm{No}$, {$\mathrm{Lr}$}, and {$\mathrm{Rf}$}},\ }\href
  {https://doi.org/10.1103/PhysRevA.90.012504} {\bibfield  {journal} {\bibinfo
  {journal} {Physical Review A}\ }\textbf {\bibinfo {volume} {90}},\ \bibinfo
  {pages} {012504} (\bibinfo {year} {2014}{\natexlab{a}})}\BibitemShut
  {NoStop}%
\bibitem [{\citenamefont {Fritzsche}(2005)}]{fritzsche-05}%
  \BibitemOpen
  \bibfield  {author} {\bibinfo {author} {\bibfnamefont {S.}~\bibnamefont
  {Fritzsche}},\ }\bibfield  {title} {\bibinfo {title} {On the accuracy of
  valence--shell computations for heavy and super--heavy elements},\ }\href
  {https://doi.org/10.1140/epjd/e2005-00013-1} {\bibfield  {journal} {\bibinfo
  {journal} {The European Physical Journal D-Atomic, Molecular, Optical and
  Plasma Physics}\ }\textbf {\bibinfo {volume} {33}},\ \bibinfo {pages} {15}
  (\bibinfo {year} {2005})}\BibitemShut {NoStop}%
\bibitem [{nis(2013)}]{nist}%
  \BibitemOpen
  \href@noop {} {\bibinfo {title} {Nist atomic spectroscopic database}},\
  \bibinfo {howpublished}
  {\url{https://physics.nist.gov/PhysRefData/ASD/levels_form.html}} (\bibinfo
  {year} {2013})\BibitemShut {NoStop}%
\bibitem [{\citenamefont {Kara{\c{c}}oban}\ and\ \citenamefont
  {{\"O}zdemir}(2011)}]{karaccoban-11}%
  \BibitemOpen
  \bibfield  {author} {\bibinfo {author} {\bibfnamefont {B.}~\bibnamefont
  {Kara{\c{c}}oban}}\ and\ \bibinfo {author} {\bibfnamefont {L.}~\bibnamefont
  {{\"O}zdemir}},\ }\bibfield  {title} {\bibinfo {title} {Transition energies
  of ytterbium (\rm{Z}= 70)},\ }\href {https://doi.org/10.5560/ZNA.2011-0003}
  {\bibfield  {journal} {\bibinfo  {journal} {Zeitschrift f{\"u}r
  Naturforschung A}\ }\textbf {\bibinfo {volume} {66}},\ \bibinfo {pages} {543}
  (\bibinfo {year} {2011})}\BibitemShut {NoStop}%
\bibitem [{\citenamefont {G{\'a}lvez}\ \emph {et~al.}(2008)\citenamefont
  {G{\'a}lvez}, \citenamefont {Buend{\'\i}a}, \citenamefont {Maldonado},\ and\
  \citenamefont {Sarsa}}]{galvez-08}%
  \BibitemOpen
  \bibfield  {author} {\bibinfo {author} {\bibfnamefont {F.}~\bibnamefont
  {G{\'a}lvez}}, \bibinfo {author} {\bibfnamefont {E.}~\bibnamefont
  {Buend{\'\i}a}}, \bibinfo {author} {\bibfnamefont {P.}~\bibnamefont
  {Maldonado}},\ and\ \bibinfo {author} {\bibfnamefont {A.}~\bibnamefont
  {Sarsa}},\ }\bibfield  {title} {\bibinfo {title} {Optimized effective
  potential energies and ionization potentials for the atoms \rm{Li} to
  \rm{Ra}},\ }\href {https://doi.org/10.1140/epjd/e2008-00222-0} {\bibfield
  {journal} {\bibinfo  {journal} {The European Physical Journal D}\ }\textbf
  {\bibinfo {volume} {50}},\ \bibinfo {pages} {229} (\bibinfo {year}
  {2008})}\BibitemShut {NoStop}%
\bibitem [{\citenamefont {Migdalek}\ and\ \citenamefont
  {Baylis}(1986)}]{migdalek-86}%
  \BibitemOpen
  \bibfield  {author} {\bibinfo {author} {\bibfnamefont {J.}~\bibnamefont
  {Migdalek}}\ and\ \bibinfo {author} {\bibfnamefont {W.}~\bibnamefont
  {Baylis}},\ }\bibfield  {title} {\bibinfo {title} {Correlation effects in a
  relativistic calculation of the 6s$^{2}$ $^{1}{S}_{0}$ - 6s6p$^{1}{P}_{1}$
  transition in ytterbium},\ }\href {https://doi.org/10.1103/PhysRevA.33.1417}
  {\bibfield  {journal} {\bibinfo  {journal} {Physical Review A}\ }\textbf
  {\bibinfo {volume} {33}},\ \bibinfo {pages} {1417} (\bibinfo {year}
  {1986})}\BibitemShut {NoStop}%
\bibitem [{\citenamefont {Sugar}(1974)}]{sugar-74}%
  \BibitemOpen
  \bibfield  {author} {\bibinfo {author} {\bibfnamefont {J.}~\bibnamefont
  {Sugar}},\ }\bibfield  {title} {\bibinfo {title} {Revised ionization energies
  of the neutral actinides},\ }\href {https://doi.org/10.1063/1.1680874}
  {\bibfield  {journal} {\bibinfo  {journal} {The Journal of Chemical Physics}\
  }\textbf {\bibinfo {volume} {60}},\ \bibinfo {pages} {4103} (\bibinfo {year}
  {1974})}\BibitemShut {NoStop}%
\bibitem [{\citenamefont {Johnson}(2007)}]{johnsson_book}%
  \BibitemOpen
  \bibfield  {author} {\bibinfo {author} {\bibfnamefont {W.~R.}\ \bibnamefont
  {Johnson}},\ }\href {https://doi.org/10.1007/978-3-540-68013-0} {\bibinfo
  {title} {Atomic structure theory. lectures on atomic physics}} (\bibinfo
  {year} {2007})\BibitemShut {NoStop}%
\bibitem [{\citenamefont {{Froese Fischer}}\ \emph {et~al.}(2019)\citenamefont
  {{Froese Fischer}}, \citenamefont {Gaigalas}, \citenamefont {Jönsson},\ and\
  \citenamefont {Bieroń}}]{fischer-19}%
  \BibitemOpen
  \bibfield  {author} {\bibinfo {author} {\bibfnamefont {C.}~\bibnamefont
  {{Froese Fischer}}}, \bibinfo {author} {\bibfnamefont {G.}~\bibnamefont
  {Gaigalas}}, \bibinfo {author} {\bibfnamefont {P.}~\bibnamefont {Jönsson}},\
  and\ \bibinfo {author} {\bibfnamefont {J.}~\bibnamefont {Bieroń}},\
  }\bibfield  {title} {\bibinfo {title} {Grasp2018—a fortran 95 version of
  the general relativistic atomic structure package},\ }\href
  {https://doi.org/https://doi.org/10.1016/j.cpc.2018.10.032} {\bibfield
  {journal} {\bibinfo  {journal} {Computer Physics Communications}\ }\textbf
  {\bibinfo {volume} {237}},\ \bibinfo {pages} {184} (\bibinfo {year}
  {2019})}\BibitemShut {NoStop}%
\bibitem [{\citenamefont {Ekman}\ \emph {et~al.}(2019)\citenamefont {Ekman},
  \citenamefont {J{\"o}nsson}, \citenamefont {Godefroid}, \citenamefont
  {Naz{\'e}}, \citenamefont {Gaigalas},\ and\ \citenamefont
  {Biero{\'n}}}]{ekman-19}%
  \BibitemOpen
  \bibfield  {author} {\bibinfo {author} {\bibfnamefont {J.}~\bibnamefont
  {Ekman}}, \bibinfo {author} {\bibfnamefont {P.}~\bibnamefont {J{\"o}nsson}},
  \bibinfo {author} {\bibfnamefont {M.}~\bibnamefont {Godefroid}}, \bibinfo
  {author} {\bibfnamefont {C.}~\bibnamefont {Naz{\'e}}}, \bibinfo {author}
  {\bibfnamefont {G.}~\bibnamefont {Gaigalas}},\ and\ \bibinfo {author}
  {\bibfnamefont {J.}~\bibnamefont {Biero{\'n}}},\ }\bibfield  {title}
  {\bibinfo {title} {Ris 4: A program for relativistic isotope shift
  calculations},\ }\href {https://doi.org/10.1016/j.cpc.2018.08.017} {\bibfield
   {journal} {\bibinfo  {journal} {Computer Physics Communications}\ }\textbf
  {\bibinfo {volume} {235}},\ \bibinfo {pages} {433} (\bibinfo {year}
  {2019})}\BibitemShut {NoStop}%
\bibitem [{\citenamefont {Sahoo}\ and\ \citenamefont {Das}(2008)}]{sahoo-08}%
  \BibitemOpen
  \bibfield  {author} {\bibinfo {author} {\bibfnamefont {B.~K.}\ \bibnamefont
  {Sahoo}}\ and\ \bibinfo {author} {\bibfnamefont {B.~P.}\ \bibnamefont
  {Das}},\ }\bibfield  {title} {\bibinfo {title} {Relativistic coupled-cluster
  studies of dipole polarizabilities in closed-shell atoms},\ }\href
  {https://doi.org/10.1103/PhysRevA.77.062516} {\bibfield  {journal} {\bibinfo
  {journal} {Phys. Rev. A}\ }\textbf {\bibinfo {volume} {77}},\ \bibinfo
  {pages} {062516} (\bibinfo {year} {2008})}\BibitemShut {NoStop}%
\bibitem [{\citenamefont {Thierfelder}\ and\ \citenamefont
  {Schwerdtfeger}(2009{\natexlab{b}})}]{peter-09}%
  \BibitemOpen
  \bibfield  {author} {\bibinfo {author} {\bibfnamefont {C.}~\bibnamefont
  {Thierfelder}}\ and\ \bibinfo {author} {\bibfnamefont {P.}~\bibnamefont
  {Schwerdtfeger}},\ }\bibfield  {title} {\bibinfo {title} {Effect of
  relativity and electron correlation in static dipole polarizabilities of
  ytterbium and nobelium},\ }\href {https://doi.org/10.1103/PhysRevA.79.032512}
  {\bibfield  {journal} {\bibinfo  {journal} {Phys. Rev. A}\ }\textbf {\bibinfo
  {volume} {79}},\ \bibinfo {pages} {032512} (\bibinfo {year}
  {2009}{\natexlab{b}})}\BibitemShut {NoStop}%
\bibitem [{\citenamefont {Dzuba}\ and\ \citenamefont
  {Derevianko}(2010)}]{dzuba-10}%
  \BibitemOpen
  \bibfield  {author} {\bibinfo {author} {\bibfnamefont {V.~A.}\ \bibnamefont
  {Dzuba}}\ and\ \bibinfo {author} {\bibfnamefont {A.}~\bibnamefont
  {Derevianko}},\ }\bibfield  {title} {\bibinfo {title} {Dynamic
  polarizabilities and related properties of clock states of the ytterbium
  atom},\ }\href {https://doi.org/10.1088/0953-4075/43/7/074011} {\bibfield
  {journal} {\bibinfo  {journal} {Journal of Physics B: Atomic, Molecular and
  Optical Physics}\ }\textbf {\bibinfo {volume} {43}},\ \bibinfo {pages}
  {074011} (\bibinfo {year} {2010})}\BibitemShut {NoStop}%
\bibitem [{\citenamefont {{Buchachenko, A. A.}}(2011)}]{buchachenko-11}%
  \BibitemOpen
  \bibfield  {author} {\bibinfo {author} {\bibnamefont {{Buchachenko, A.
  A.}}},\ }\bibfield  {title} {\bibinfo {title} {Ab~initio dipole
  polarizabilities and quadrupole moments of the lowest excited states of
  atomic {Yb}},\ }\href {https://doi.org/10.1140/epjd/e2010-10413-7} {\bibfield
   {journal} {\bibinfo  {journal} {Eur. Phys. J. D}\ }\textbf {\bibinfo
  {volume} {61}},\ \bibinfo {pages} {291} (\bibinfo {year} {2011})}\BibitemShut
  {NoStop}%
\bibitem [{\citenamefont {Dzuba}\ \emph
  {et~al.}(2014{\natexlab{b}})\citenamefont {Dzuba}, \citenamefont {Kozlov},\
  and\ \citenamefont {Flambaum}}]{dzuba-14a}%
  \BibitemOpen
  \bibfield  {author} {\bibinfo {author} {\bibfnamefont {V.~A.}\ \bibnamefont
  {Dzuba}}, \bibinfo {author} {\bibfnamefont {A.}~\bibnamefont {Kozlov}},\ and\
  \bibinfo {author} {\bibfnamefont {V.~V.}\ \bibnamefont {Flambaum}},\
  }\bibfield  {title} {\bibinfo {title} {Scalar static polarizabilities of
  lanthanides and actinides},\ }\href
  {https://doi.org/10.1103/PhysRevA.89.042507} {\bibfield  {journal} {\bibinfo
  {journal} {Phys. Rev. A}\ }\textbf {\bibinfo {volume} {89}},\ \bibinfo
  {pages} {042507} (\bibinfo {year} {2014}{\natexlab{b}})}\BibitemShut
  {NoStop}%
\bibitem [{\citenamefont {Dzuba}(2016)}]{dzuba-16}%
  \BibitemOpen
  \bibfield  {author} {\bibinfo {author} {\bibfnamefont {V.~A.}\ \bibnamefont
  {Dzuba}},\ }\bibfield  {title} {\bibinfo {title} {{Ionization potentials and
  polarizabilities of superheavy elements from Db to Cn (Z=105--112)}},\ }\href
  {https://doi.org/10.1103/PhysRevA.93.032519} {\bibfield  {journal} {\bibinfo
  {journal} {Phys. Rev. A}\ }\textbf {\bibinfo {volume} {93}},\ \bibinfo
  {pages} {032519} (\bibinfo {year} {2016})}\BibitemShut {NoStop}%
\bibitem [{\citenamefont {Zhang}\ \emph {et~al.}(2009)\citenamefont {Zhang},
  \citenamefont {Dalgarno},\ and\ \citenamefont {C\^ot\'e}}]{zhang-09}%
  \BibitemOpen
  \bibfield  {author} {\bibinfo {author} {\bibfnamefont {P.}~\bibnamefont
  {Zhang}}, \bibinfo {author} {\bibfnamefont {A.}~\bibnamefont {Dalgarno}},\
  and\ \bibinfo {author} {\bibfnamefont {R.}~\bibnamefont {C\^ot\'e}},\
  }\bibfield  {title} {\bibinfo {title} {Scattering of {Yb and
  ${\text{Yb}}^{+}$}},\ }\href {https://doi.org/10.1103/PhysRevA.80.030703}
  {\bibfield  {journal} {\bibinfo  {journal} {Phys. Rev. A}\ }\textbf {\bibinfo
  {volume} {80}},\ \bibinfo {pages} {030703} (\bibinfo {year}
  {2009})}\BibitemShut {NoStop}%
\bibitem [{\citenamefont {Safronova}\ \emph {et~al.}(2012)\citenamefont
  {Safronova}, \citenamefont {Porsev},\ and\ \citenamefont
  {Clark}}]{safronova-12}%
  \BibitemOpen
  \bibfield  {author} {\bibinfo {author} {\bibfnamefont {M.~S.}\ \bibnamefont
  {Safronova}}, \bibinfo {author} {\bibfnamefont {S.~G.}\ \bibnamefont
  {Porsev}},\ and\ \bibinfo {author} {\bibfnamefont {C.~W.}\ \bibnamefont
  {Clark}},\ }\bibfield  {title} {\bibinfo {title} {Ytterbium in quantum gases
  and atomic clocks: van der waals interactions and blackbody shifts},\ }\href
  {https://doi.org/10.1103/PhysRevLett.109.230802} {\bibfield  {journal}
  {\bibinfo  {journal} {Phys. Rev. Lett.}\ }\textbf {\bibinfo {volume} {109}},\
  \bibinfo {pages} {230802} (\bibinfo {year} {2012})}\BibitemShut {NoStop}%
\bibitem [{\citenamefont {Wang}\ and\ \citenamefont {Dolg}(1998)}]{wang-98}%
  \BibitemOpen
  \bibfield  {author} {\bibinfo {author} {\bibfnamefont {Y.}~\bibnamefont
  {Wang}}\ and\ \bibinfo {author} {\bibfnamefont {M.}~\bibnamefont {Dolg}},\
  }\bibfield  {title} {\bibinfo {title} {Pseudopotential study of the ground
  and excited states of {Yb$_2$}},\ }\href
  {https://doi.org/10.1007/s002140050373} {\bibfield  {journal} {\bibinfo
  {journal} {Theoretical Chemistry Accounts}\ }\textbf {\bibinfo {volume}
  {100}},\ \bibinfo {pages} {124} (\bibinfo {year} {1998})}\BibitemShut
  {NoStop}%
\bibitem [{\citenamefont {Yoshizawa}\ \emph {et~al.}(2016)\citenamefont
  {Yoshizawa}, \citenamefont {Zou},\ and\ \citenamefont
  {Cremer}}]{yoshizawa-16}%
  \BibitemOpen
  \bibfield  {author} {\bibinfo {author} {\bibfnamefont {T.}~\bibnamefont
  {Yoshizawa}}, \bibinfo {author} {\bibfnamefont {W.}~\bibnamefont {Zou}},\
  and\ \bibinfo {author} {\bibfnamefont {D.}~\bibnamefont {Cremer}},\
  }\bibfield  {title} {\bibinfo {title} {Calculations of electric dipole
  moments and static dipole polarizabilities based on the two-component
  normalized elimination of the small component method},\ }\href
  {https://doi.org/10.1063/1.4964765} {\bibfield  {journal} {\bibinfo
  {journal} {The Journal of Chemical Physics}\ }\textbf {\bibinfo {volume}
  {145}},\ \bibinfo {pages} {184104} (\bibinfo {year} {2016})},\ \Eprint
  {https://arxiv.org/abs/https://doi.org/10.1063/1.4964765}
  {https://doi.org/10.1063/1.4964765} \BibitemShut {NoStop}%
\bibitem [{\citenamefont {Buchachenko}\ \emph {et~al.}(2006)\citenamefont
  {Buchachenko}, \citenamefont {Szczesniak},\ and\ \citenamefont
  {Chalasinski}}]{buchachenko-06}%
  \BibitemOpen
  \bibfield  {author} {\bibinfo {author} {\bibfnamefont {A.~A.}\ \bibnamefont
  {Buchachenko}}, \bibinfo {author} {\bibfnamefont {M.~M.}\ \bibnamefont
  {Szczesniak}},\ and\ \bibinfo {author} {\bibfnamefont {G.}~\bibnamefont
  {Chalasinski}},\ }\bibfield  {title} {\bibinfo {title} {van der waals
  interactions and dipole polarizabilities of lanthanides: {Tm(F2)–He and
  Yb(S1)–He potentials}},\ }\href {https://doi.org/10.1063/1.2176602}
  {\bibfield  {journal} {\bibinfo  {journal} {The Journal of Chemical Physics}\
  }\textbf {\bibinfo {volume} {124}},\ \bibinfo {pages} {114301} (\bibinfo
  {year} {2006})},\ \Eprint
  {https://arxiv.org/abs/https://doi.org/10.1063/1.2176602}
  {https://doi.org/10.1063/1.2176602} \BibitemShut {NoStop}%
\bibitem [{\citenamefont {Zhang}\ and\ \citenamefont
  {Dalgarno}(2007)}]{zhang-07}%
  \BibitemOpen
  \bibfield  {author} {\bibinfo {author} {\bibfnamefont {P.}~\bibnamefont
  {Zhang}}\ and\ \bibinfo {author} {\bibfnamefont {A.}~\bibnamefont
  {Dalgarno}},\ }\bibfield  {title} {\bibinfo {title} {Static dipole
  polarizability of ytterbium},\ }\href {https://doi.org/10.1021/jp0750856}
  {\bibfield  {journal} {\bibinfo  {journal} {The Journal of Physical Chemistry
  A}\ }\textbf {\bibinfo {volume} {111}},\ \bibinfo {pages} {12471} (\bibinfo
  {year} {2007})},\ \bibinfo {note} {pMID: 17915845},\ \Eprint
  {https://arxiv.org/abs/https://doi.org/10.1021/jp0750856}
  {https://doi.org/10.1021/jp0750856} \BibitemShut {NoStop}%
\bibitem [{\citenamefont {Chu}\ \emph {et~al.}(2007)\citenamefont {Chu},
  \citenamefont {Dalgarno},\ and\ \citenamefont {Groenenboom}}]{xchu-07}%
  \BibitemOpen
  \bibfield  {author} {\bibinfo {author} {\bibfnamefont {X.}~\bibnamefont
  {Chu}}, \bibinfo {author} {\bibfnamefont {A.}~\bibnamefont {Dalgarno}},\ and\
  \bibinfo {author} {\bibfnamefont {G.~C.}\ \bibnamefont {Groenenboom}},\
  }\bibfield  {title} {\bibinfo {title} {Dynamic polarizabilities of
  rare-earth-metal atoms and dispersion coefficients for their interaction with
  helium atoms},\ }\href {https://doi.org/10.1103/PhysRevA.75.032723}
  {\bibfield  {journal} {\bibinfo  {journal} {Phys. Rev. A}\ }\textbf {\bibinfo
  {volume} {75}},\ \bibinfo {pages} {032723} (\bibinfo {year}
  {2007})}\BibitemShut {NoStop}%
\bibitem [{\citenamefont {Buchachenko}\ \emph {et~al.}(2007)\citenamefont
  {Buchachenko}, \citenamefont {Cha{\l}asi{\'{n}}ski},\ and\ \citenamefont
  {Szcze{\'{s}}niak}}]{buchachenko-07}%
  \BibitemOpen
  \bibfield  {author} {\bibinfo {author} {\bibfnamefont {A.~A.}\ \bibnamefont
  {Buchachenko}}, \bibinfo {author} {\bibfnamefont {G.}~\bibnamefont
  {Cha{\l}asi{\'{n}}ski}},\ and\ \bibinfo {author} {\bibfnamefont {M.~M.}\
  \bibnamefont {Szcze{\'{s}}niak}},\ }\bibfield  {title} {\bibinfo {title}
  {Diffuse basis functions for small-core relativistic pseudopotential basis
  sets and static dipole polarizabilities of selected lanthanides {La, Sm, Eu,
  Tm and Yb}},\ }\href {https://doi.org/10.1007/s11224-007-9243-1} {\bibfield
  {journal} {\bibinfo  {journal} {Structural Chemistry}\ }\textbf {\bibinfo
  {volume} {18}},\ \bibinfo {pages} {769} (\bibinfo {year} {2007})}\BibitemShut
  {NoStop}%
\bibitem [{\citenamefont {Sahoo}\ and\ \citenamefont {Das}(2018)}]{sahoo-18}%
  \BibitemOpen
  \bibfield  {author} {\bibinfo {author} {\bibfnamefont {B.~K.}\ \bibnamefont
  {Sahoo}}\ and\ \bibinfo {author} {\bibfnamefont {B.~P.}\ \bibnamefont
  {Das}},\ }\bibfield  {title} {\bibinfo {title} {The role of relativistic
  many-body theory in probing new physics beyond the standard model via the
  electric dipole moments of diamagnetic atoms},\ }\href
  {https://doi.org/10.1088/1742-6596/1041/1/012014} {\bibfield  {journal}
  {\bibinfo  {journal} {Journal of Physics: Conference Series}\ }\textbf
  {\bibinfo {volume} {1041}},\ \bibinfo {pages} {012014} (\bibinfo {year}
  {2018})}\BibitemShut {NoStop}%
\bibitem [{\citenamefont {Ma}\ \emph {et~al.}(2015)\citenamefont {Ma},
  \citenamefont {Indergaard}, \citenamefont {Zhang}, \citenamefont {Larkin},
  \citenamefont {Moro},\ and\ \citenamefont {de~Heer}}]{lei-15}%
  \BibitemOpen
  \bibfield  {author} {\bibinfo {author} {\bibfnamefont {L.}~\bibnamefont
  {Ma}}, \bibinfo {author} {\bibfnamefont {J.}~\bibnamefont {Indergaard}},
  \bibinfo {author} {\bibfnamefont {B.}~\bibnamefont {Zhang}}, \bibinfo
  {author} {\bibfnamefont {I.}~\bibnamefont {Larkin}}, \bibinfo {author}
  {\bibfnamefont {R.}~\bibnamefont {Moro}},\ and\ \bibinfo {author}
  {\bibfnamefont {W.~A.}\ \bibnamefont {de~Heer}},\ }\bibfield  {title}
  {\bibinfo {title} {Measured atomic ground-state polarizabilities of 35
  metallic elements},\ }\href {https://doi.org/10.1103/PhysRevA.91.010501}
  {\bibfield  {journal} {\bibinfo  {journal} {Phys. Rev. A}\ }\textbf {\bibinfo
  {volume} {91}},\ \bibinfo {pages} {010501} (\bibinfo {year}
  {2015})}\BibitemShut {NoStop}%
\bibitem [{\citenamefont {Beloy}(2012)}]{beloy-12}%
  \BibitemOpen
  \bibfield  {author} {\bibinfo {author} {\bibfnamefont {K.}~\bibnamefont
  {Beloy}},\ }\bibfield  {title} {\bibinfo {title} {Experimental constraints on
  the polarizabilities of the {$6s^2$ $^1S_{0}$ and $6s6p$ $^3P_0^o$ states of
  Yb}},\ }\href {https://doi.org/10.1103/PhysRevA.86.022521} {\bibfield
  {journal} {\bibinfo  {journal} {Phys. Rev. A}\ }\textbf {\bibinfo {volume}
  {86}},\ \bibinfo {pages} {022521} (\bibinfo {year} {2012})}\BibitemShut
  {NoStop}%
\bibitem [{\citenamefont {Martins}\ \emph {et~al.}(2016)\citenamefont
  {Martins}, \citenamefont {Jorge}, \citenamefont {Franco},\ and\ \citenamefont
  {Ferreira}}]{martin-16}%
  \BibitemOpen
  \bibfield  {author} {\bibinfo {author} {\bibfnamefont {L.~S.~C.}\
  \bibnamefont {Martins}}, \bibinfo {author} {\bibfnamefont {F.~E.}\
  \bibnamefont {Jorge}}, \bibinfo {author} {\bibfnamefont {M.~L.}\ \bibnamefont
  {Franco}},\ and\ \bibinfo {author} {\bibfnamefont {I.~B.}\ \bibnamefont
  {Ferreira}},\ }\bibfield  {title} {\bibinfo {title} {All-electron gaussian
  basis sets of double zeta quality for the actinides},\ }\href
  {https://doi.org/10.1063/1.4973377} {\bibfield  {journal} {\bibinfo
  {journal} {The Journal of Chemical Physics}\ }\textbf {\bibinfo {volume}
  {145}},\ \bibinfo {pages} {244113} (\bibinfo {year} {2016})},\ \Eprint
  {https://arxiv.org/abs/https://doi.org/10.1063/1.4973377}
  {https://doi.org/10.1063/1.4973377} \BibitemShut {NoStop}%
\bibitem [{\citenamefont {Zatsarinny}\ and\ \citenamefont {{Froese
  Fischer}}(2016)}]{zatsarinny-16}%
  \BibitemOpen
  \bibfield  {author} {\bibinfo {author} {\bibfnamefont {O.}~\bibnamefont
  {Zatsarinny}}\ and\ \bibinfo {author} {\bibfnamefont {C.}~\bibnamefont
  {{Froese Fischer}}},\ }\bibfield  {title} {\bibinfo {title} {{DBSR-HF: A
  B-spline Dirac–Hartree–Fock program}},\ }\href
  {https://doi.org/https://doi.org/10.1016/j.cpc.2015.12.023} {\bibfield
  {journal} {\bibinfo  {journal} {Computer Physics Communications}\ }\textbf
  {\bibinfo {volume} {202}},\ \bibinfo {pages} {287 } (\bibinfo {year}
  {2016})}\BibitemShut {NoStop}%
\end{thebibliography}%

\end{document}